\documentclass[aps,pre,floatfix,twocolumn,showpacs]{revtex4}
\usepackage{graphicx}
\usepackage{amsmath}
\usepackage{verbatim}
\usepackage{bm}

\begin{document}

\title{Frequency and temperature dependence of the anomalous ac Hall conductivity in a chiral \boldmath $p_x+ip_y$ superconductor with impurities}
\author{Roman M. Lutchyn}
\author{Pavel Nagornykh}
\author{Victor M.~Yakovenko}
\affiliation{Joint Quantum Institute, Department of Physics,
University of Maryland, College Park, MD 20742-4111, USA}

\date{
 compiled \today}

\begin{abstract}
We calculate frequency and temperature dependence of the anomalous ac Hall conductivity induced by impurity scattering in a chiral $p_x+ip_y$ superconductor, such as $\rm Sr_2RuO_4$, with spontaneous time-reversal-symmetry breaking in the absence of an external magnetic field.  We consider two models of disorder, Gaussian and non-Gaussian, characterized by the second and third moments of the random impurity potential, respectively.  Within both models, we find that the anomalous Hall conductivity has a finite real value at zero frequency, exhibits singularities at the threshold of photon absorption across the superconducting gap, and decays as some power of the high frequency $\omega$.  The Hall conductivity increases linearly with the decrease of temperature below the superconducting transition and saturates at zero temperature.  Using our results for the high-frequency Hall conductivity, we estimate the polar Kerr angle for light reflection from the material and compare it with the experimental measurements in $\rm Sr_2RuO_4$ by Xia \emph{et al.}, Phys. Rev. Lett. {\bf 97}, 167002 (2006).
\end{abstract}

\maketitle

\section{Introduction}
\label{Sec:Intro}

Recent development of an optical apparatus based on the Sagnac interferometer by the group of Kapitulnik led to spectacular discoveries of the spontaneous time-reversal symmetry breaking (TRSB) in the unconventional superconductors $\rm Sr_2RuO_4$ \cite{xia_prl'06} and $\rm YBa_2Cu_3O_{6+x}$ \cite{Xia_PRL'08}.  These experiments measured the rotation angle $\theta_K$ of the polarization plane of light upon normal reflected from the material surface.  A non-zero value of the Kerr angle $\theta_K$ indicates the TRSB in the sample \cite{Kapitulnik_NJP'09}.  Positive and negative values of $\theta_K$ represent clockwise and counterclockwise rotation of polarization and indicate two possible ways of spontaneous breaking of the time-reversal symmetry.  In $\rm Sr_2RuO_4$, a non-zero $\theta_K$ appears at the superconducting transition temperature  $T_c=1.5$K, so the TRSB in this material is clearly related to the onset of superconductivity  \cite{xia_prl'06}.  In contrast, in the underdoped $\rm YBa_2Cu_3O_{6+x}$, a non-zero $\theta_K$ appears in the temperature range where the pseudogap develops, which is much higher than the superconducting transition temperature \cite{Xia_PRL'08}.  Thus, the TRSB in cuprates appears to be due to a second-order phase transition with an order parameter unrelated to superconductivity.

In this paper, we focus on manifestations of the TRSB in $\rm Sr_2RuO_4$, which is a quasi-two-dimensional (Q2D) metal consisting of weakly coupled conducting layers in the $(x,y)$ plane \cite{Maeno_Nature'94,Mackenzie_RevModPhys'03}.  It was proposed theoretically that the superconducting pairing in $\rm Sr_2RuO_4$ is spin-triplet with the chiral $p_x\pm ip_y$ orbital symmetry \cite{Rice95}.  In this state, the Cooper pairs have the angular momentum $L_z=\pm1$ pointing perpendicular to the conducting layers of $\rm Sr_2RuO_4$.  This order parameter breaks the time-reversal symmetry and is the two-dimensional analogue of the A phase in superfluid $^3$He~\cite{Volovik_JETP'88,Mineev-book,maeno_PhysToday'01}.  There is substantial experimental evidence indicating that $\rm Sr_2RuO_4$ is indeed a spin-triplet~\cite{Ishida_Nature'98,duffy_prl'00} $p$-wave~\cite{Nelson_science'04} superconductor.  However, the chiral orbital symmetry of the order parameter and the violation of the time-reversal symmetry are still under debate~\cite{Kallin_JPCM'09}.  The early evidence for the TRSB came from the muon spin-relaxation measurements~\cite{Luke_Nature'98}.  Recent observations of the spontaneous polar Kerr effect in optics~\cite{xia_prl'06} and dynamical domains of the superconducting order parameter in Josephson junctions~\cite{Kidwingira_science'06} give strong evidence for the chiral pairing with the TRSB~\cite{day_PhysToday'06,Rice_science'06}.  However, the scanning SQUID and Hall probe experiments~\cite{Moller_prb'05,kirtley_prb'07}, designed to search for domains with opposite chiralities on the surface of $\rm Sr_2RuO_4$, did not find any evidence for such domains and for the TRSB.  The discrepancy between these experimental results shows that the TRSB in superconductors is not fully understood and requires further investigation~\cite{Kallin_JPCM'09}.  Chiral superconductors with the TRSB are expected to have many unusual properties, e.g., certain vortex excitations with zero-energy Majorana modes in the core~\cite{read_prb'00, ivanov_prl'01}.  These non-local quasiparticle excitations obey non-Abelian statistics and are studied in the context of topological quantum computation~\cite{nayak_RevModPhys'08}.  The fractional quantum Hall state at the filling factor 5/2 was proposed to be analogous to the $p_x+ip_y$ superconductivity \cite{read_prb'00,Read_PRB'09}.

Here we study an important manifestation of the TRSB in the superconducting state: the emergence of anomalous (spontaneous) Hall conductivity $\sigma_{xy}$ in the absence of an external magnetic field.  This study is motivated by the experimental measurement~\cite{xia_prl'06} of the polar Kerr effect in $\rm Sr_2RuO_4$.  According to the textbook calculations using classical electrodynamics~\cite{White-Geballe}, the Kerr angle $\theta_K$ is determined by the ac Hall conductivity $\sigma_{xy}(\omega)$, where $\omega$ is the frequency of light in the experiment.  Thus, the experimental observation~\cite{xia_prl'06} of a non-zero $\theta_K$ at $T<T_c$ signifies appearance of a non-zero anomalous ac Hall conductivity in the material.  However, a theoretical calculation of the anomalous ac Hall conductivity for a chiral $p_x+ip_y$ superconductor turned out to be a rather non-trivial problem and led to some controversies briefly reviewed in the next Subsection.

\subsection{Theories of the anomalous ac Hall conductivity in a clean chiral \boldmath $p_x+ip_y$ superconductor}
\label{Sec:Theories}

Most of the previous theoretical calculations of the anomalous Hall
conductivity focused on the clean limit for a chiral
translationally-invariant superconductor in the absence of
impurities.  However, a straightforward calculation of the
current-current correlation function shown in Fig.~\ref{fig:no_line}
and involving interaction with the $A_x$ and $A_y$ components of the
electromagnetic vector potential gives zero result
\cite{Volovik_JETP'88,Joynt_PRB'91,Goryo_PhysLettA'98,Goryo_PhysLettA'99,Horovitz_epl'02,Horovitz_PRB'03,Stone_PRB'04,yakovenko_prl'07,lutchyn_prb'08}.
This implies that the Hall conductivity $\sigma_{xy}$ in a clean
chiral superconductor is zero, even though a non-zero value is
nominally permitted by the TRSB.  Kim \textit{et
al.}~\cite{kim_prl'08} obtained a non-zero value of $\sigma_{xy}$
for an anisotropic superconducting pairing, but their calculated
conductivity tensor is symmetric $\sigma_{xy}=\sigma_{yx}$. So, it
does not represent the antisymmetric Hall conductivity
($\sigma_{xy}=-\sigma_{yx}$) and does not contribute to the polar
Kerr effect.

On the other hand, it was found that the charge-current correlation
function, represented by the Feynman diagram with the scalar
electromagnetic potential $A_0$ and the vector potential $A_x$ or
$A_y$, is non-zero in chiral superconductors
\cite{Volovik_JETP'88,Goryo_PhysLettA'98,Goryo_PhysLettA'99,Horovitz_epl'02,Horovitz_PRB'03,Stone_PRB'04,yakovenko_prl'07,lutchyn_prb'08}.
It describes the magneto-electric effect, i.e.,\ a change in the
electric charge density in response to an applied magnetic field
$B_z$.  As a consequence of this effect, a magnetic vortex would
acquire the fractional electric charge $e/4$ \cite{Goryo_PRB'00}.  A
setup for an experimental detection of the magneto-electric effect
in chiral superconductors was proposed in
Ref.~\cite{lutchyn_prb'08}.

In conventional (non-superconducting) quantum Hall systems, the
magneto-electric effect is directly related to the Hall conductivity
via the \u{S}treda formula~\cite{Streda}.  However, in
superconductors, this relation is invalid, and the magneto-electric
effect does not imply the Hall effect.  Incorrect claims were made
in Refs.~\cite{Volovik_JETP'88,yakovenko_prl'07,mineev_prb'07} on
the basis of the magneto-electric effect that clean chiral
superconductors have a non-zero anomalous Hall effect.  It was shown
in
Refs.~\cite{Goryo_PhysLettA'98,Goryo_PhysLettA'99,Horovitz_epl'02,Horovitz_PRB'03,lutchyn_prb'08,roy_prb'08}
that, when the contribution from the collective current of the
superconducting condensate is properly taken into account in a
gauge-invariant manner, the Hall conductivity vanishes.  This result
is also in qualitative agreement with the conclusions of
Ref.~\cite{Furusaki_PRB'01}.  In principle, a non-zero value of the
Hall conductivity $\sigma_{xy}(\bm q,\omega)$ can be obtained for a
non-zero in-plane wave vector $\bm q$
\cite{Goryo_PhysLettA'98,Goryo_PhysLettA'99,Horovitz_epl'02,Horovitz_PRB'03,lutchyn_prb'08}.
However, the estimates done in Ref.~\cite{lutchyn_prb'08} using $\bm
q$ derived from the laser beam diameter in the experiment
\cite{xia_prl'06} give the value of $\sigma_{xy}$ many orders of
magnitude smaller than what is necessary to explain the experiment.
A non-zero, but very small Kerr angle was obtained in
Ref.~\cite{Yip_JLTP'92} by considering the orbital collective modes
in a $p_x+ip_y$ superconductor.  However, the estimated value of
$\theta_K$ is too small to explain the experiment \cite{xia_prl'06}.

Refs.~\cite{Balatskii_JETP'85,mineev_prb'08} claimed that, in
unconventional superconductors, it is necessary to make the
Peierls-Onsager-style substitution of the momentum $\bm p\to\bm
p-e\bm A$ in the momentum-dependent superconducting order parameter
$\Delta(\bm p)$.  As a result, a new vertex of interaction with the
electromagnetic field $\bm A$ would appear in the Hamiltonian of the
superconducting system, and this vertex may give a non-zero value to
$\sigma_{xy}$.  However, this substitution is wrong for a number of
reasons, as explained in detail in Appendix~\ref{App:substitution}.
The absence of such substitution was shown in several papers,
including
Refs.~\cite{Volovik_JETP'88,Read_PRB'09,Vafek_PRB'01,Lee_RMP'06} and
was recently recognized in Ref.~\cite{mineev'09}.

In fact, the vanishing of the anomalous Hall conductivity for a
clean superconductor can be understood on general grounds using
Galilean invariance of the system~\cite{read_prb'00}.  In the
absence of an external magnetic field, the electric field $E_x$
applied in the $x$ direction cannot induce a center-of-mass motion
of the electron gas in the transverse $y$ direction no matter
whether the pairing between electrons is chiral or not, because
there is no total external Lorentz force acting in the $y$
direction.  Thus, the experimental observation of the polar Kerr
effect \cite{xia_prl'06} cannot be explained within any model of a
clean Galilean-invariant chiral superconductor~\cite{edges}.

The general argument given above uses the Galilean invariance of a
translationally-invariant system.  However, electrons in a crystal
are subject to a periodic lattice potential, which breaks
translational symmetry and, thus, may invalidate the general
argument.  In fact, there are well-known examples of periodic
systems, such as the TRSB topological insulators and metals, where a
topologically non-trivial band structure produces the anomalous Hall
effect \cite{Sinitsyn_JPCM'08,Haldane_PRL'04}.  However, this effect
has nothing to do with superconductivity and, thus, is not relevant
for the Kerr effect in $\rm Sr_2RuO_4$.  Relation between chiral
superconductors and the TRSB topological insulators and metals is
discussed in more detail in Appendix \ref{Sec:Insulators}.

\subsection{The role of impurities in the anomalous Hall effect}
\label{Sec:Impurities}

Following the general argument presented in the preceding
Subsection, we realize that, in order to obtain a non-zero anomalous
Hall conductivity, it is necessary to identify a physical mechanism
for producing an external force on the electron gas in the
transverse $y$ direction.  Scattering on impurities breaks
translational invariance and provides a mechanism for momentum
transfer between electrons and the crystal lattice, which may
generate a net force in the $y$ direction, if the pairing between
electrons is chiral.  Making an analogy between a Cooper pair with
the angular momentum $L_z=1$ and a spinning baseball, we observe
that the baseball does not deflect sideways when flying in vacuum,
but does deflect in the air, because friction with the air generates
the transverse Magnus force.  Similarly, scattering on impurities
can provide an effective friction between the spinning Cooper pairs
and the crystal lattice and, thus, generate a transverse force.  It
should be emphasized that a periodic lattice potential alone, in the
absence of impurities, does not produce a non-zero Hall conductivity
for a chiral $p$-wave superconductor~\cite{lutchyn_prb'08}.  It is
the random scattering on impurities which provides a mechanism for
momentum relaxation in the electron gas.

The effect of disorder on the anomalous Hall conductivity in a
$p_x+ip_y$ superconductor was studied by Goryo~\cite{goryo_prb'08}.
He considered the skew-scattering processes, which involve triple
scattering events on a given impurity.  This process requires
existence of a non-zero third moment of the impurity potential
averaged over random realizations.  Since the third moment is zero
for the commonly used Gaussian distribution, we call this type of
distribution the non-Gaussian disorder.  A similar model of disorder
was used in the context of the anomalous Hall
effect~\cite{Karplus_prb'54,
smit_physica'55,nozieres_JdP'73,sinitsyn_prb'07}, as well as the
spin Hall effect~\cite{dyakonov_JETP'71,Engel_PRL'05,tse_PRL'06},
for non-superconducting materials.  In fact,
Sinitsyn~\cite{Sinitsyn_JPCM'08} was the first to point out that
impurity scattering may contribute to the anomalous Hall effect in
chiral superconductors.

Goryo \cite{goryo_prb'08} calculated the anomalous ac Hall
conductivity $\sigma_{xy}(\omega)$ for a $p_x+ip_y$ superconductor
in the limit of high frequencies $\omega\gg\Delta_0$, where
$2\Delta_0$ is the superconducting gap, and for $T$ close to $T_c$.
This limit is relevant for the experiment \cite{xia_prl'06}, which
utilized $\omega=0.8$~eV, whereas the gap at $T=0$ can be estimated
from the BCS relation as $2\Delta_0=3.5\,T_c=0.46$ meV.  Goryo found
that the anomalous Hall conductivity is predominantly real in this
limit and behaves as $\sigma_{xy}^\prime(\omega) \propto
1/\omega^3$. However, this result is inconsistent with the general
causality relation
$\sigma(\omega)=\sigma^*(-\omega)$~[\onlinecite{Landau_book8}].

In this paper, we calculate the real and imaginary parts of the
antisymmetric conductivity tensor
$\sigma_{xy}(\omega)=-\sigma_{yx}(\omega)$ originating from
skew-scattering on impurities for the full range of frequency
$\omega$ and temperature $T$ consistently with the causality
relation.  We find that the factor of $i=\sqrt{-1}$ was overlooked
in the calculation of Ref.\ \cite{goryo_prb'08}.  Then we consider a
more conventional model of Gaussian disorder and calculate the
Feynman diagrams where electrons scatter on two different
impurities.  We show that these diagrams also give a non-zero
anomalous Hall conductivity for a chiral superconductor if the
particle-hole asymmetry is taken into account.  The role of the
particle-hole asymmetry was discussed in Ref.~\cite{Yip_JLTP'92} for
chiral $p$-wave superconductors and in
Refs.~\cite{Kopnin_JETP'92,Feigelman_JETPLett'95,Otterlo_PRL'95,Volovik_CJP'96}
for the Hall effect anomaly associated with vortex motion in the
high-$T_c$ superconductors.  Then we compare the expressions for
$\sigma_{xy}(\omega)$ obtained for these two models of Gaussian and
non-Gaussian disorder and discuss the dominant contribution at the
high frequencies relevant for the Kerr effect
measurements~\cite{xia_prl'06}.

The paper is organized as follows.  In Sec.~\ref{sec:II}, we first
show that $\sigma_{xy}$ vanishes for a clean chiral $p_x+ip_y$
superconductor.  Then we show that the lowest (second) order
diagrams in the strength of the impurity scattering also give a
vanishing $\sigma_{xy}$.  After that, we present a general
discussion of the higher-order contributions, which lead to a
non-zero $\sigma_{xy}$.  In Sec.~\ref{sec:skew}, we calculate
$\sigma^{(3)}_{xy}\!(\omega)$ originating from the skew-scattering
diagrams for the non-Gaussian model of disorder.  In
Sec.~\ref{sec:p-hasym}, we calculate $\sigma^{(4)}_{xy}(\omega)$ for
the Gaussian model of disorder in the presence of particle-hole
asymmetry.  The implications of our results for the experiments in
$\!\rm Sr_2RuO_4$ are discussed in Sec.~\ref{sec:PKE}, and conclusions
are given in Sec.~\ref{sec:conclusions}.  Technical details of
calculations are relegated to the Appendixes.

\section{Theoretical model and general discussion}
\label{sec:II}

In this Section, we first introduce a theoretical model for
calculation of the anomalous ac Hall conductivity.  Then, after a
brief discussion of the clean limit, we present a general discussion
of the effects of impurity scattering on the anomalous Hall
conductivity.

We use the electromagnetic gauge where the scalar potential $A_0$ is
set to zero, and the electromagnetic field is characterized by the
transverse component of the vector potential $\bm A$.  Within the
linear response approach, the current $\bm j$ appearing in response
to an infinitesimal vector potential $\bm A$ is
  \begin{align} \label{Q}
  \bm j(\bm q,\omega) = \tensor Q(\bm q,\omega) \, \bm A(\bm q,\omega),
  \end{align}
where $\tensor Q(\bm q,\omega)$ is the electromagnetic response
tensor. From this tensor, we can obtain the conductivity tensor
$\sigma_{ij}(\bm q,\omega)$ relating $\bm j$ with the electric field
$\bm E=-\partial_t\bm A$
  \begin{align} \label{eq:sigma-q}
  \sigma_{ij}(\bm q,\omega) =  -\frac{Q_{ij}(\bm q,\omega)}{i\omega}.
  \end{align}
In the experiment~\cite{xia_prl'06}, the light beam is incident
along the $z$ axis perpendicular to the two-dimensional (2D)
conducting layers of $\rm Sr_2RuO_4$, and the vector potential $\bm
A$ is polarized in the $(x,y)$ plane of the layers.  So, we
calculate the off-diagonal response function
$Q_{xy}(\omega)=-Q_{yx}(\omega)$ at $\bm q=0$, which determines the
ac Hall conductivity $\sigma_{xy}(\omega)$ and the Kerr angle
$\theta_K$ \cite{White-Geballe}.

\subsection{The model of a chiral \boldmath $p_x+ip_y$ superconductor}
\label{Sec:Hamiltonian}

The triplet superconducting pairing is characterized by the vector
$\bm d$, which determines spin polarization of the triplet Cooper
pairs~\cite{Leggett_RevModPhys'75}.  For $\rm Sr_2RuO_4$, we
consider the case where the vector $\bm d$ has a uniform,
momentum-independent orientation~\cite{maeno_PhysToday'01}.
Selecting the spin quantization axis along the vector $\bm d$, we
obtain the representation~\cite{lutchyn_prb'08} where the triplet
Cooper pairing takes place between electrons with the opposite spins
\cite{footnote2}. For the orbital symmetry, we consider the chiral
pairing potential $\Delta(\bm p)=\Delta_0(p_x\pm ip_y)/p_F$, where
$p_x$, $p_y$, and $p_F$ are the in-plane electron momenta and the
Fermi momentum.  It is convenient to write this pairing potential in
the form
\begin{equation}
  \Delta(\bm p)=\Delta_xp_x+i\Delta_yp_y
\label{Delta_xy}
\end{equation}
and set $\Delta_x=\pm\Delta_y=\Delta_0/p_F$ only at the end of the calculations.  The sign of the product
\begin{equation}
  s_{xy}\equiv{\rm sign}(\Delta_x\Delta_y)
\label{s_xy}
\end{equation}
represents the sign of the order-parameter chirality.  Notice that
the dimensionality of $\Delta_x$ and $\Delta_y$ is different from
that of $\Delta_0$.

In the Matsubara representation, the action $S_{\rm el}$ of the
electron system can be written~\cite{lutchyn_prb'08} as a $2\times2$
Nambu matrix acting on the spinor $\bm\psi(\bm
p,\varpi_l)=[\psi_\uparrow(\bm p,\varpi_l),\psi_\downarrow^\dag(-\bm
p,\varpi_l)]$, where $\psi$ and $\psi^\dag$ are the destruction and
creation operators of the electrons with the momentum $\bm p$, the
fermionic Matsubara frequency $\varpi_l$, and the spin projection
$\uparrow$ or $\downarrow$
  \begin{align} \label{S_el}
  S_{\rm el} = & \sum_{\bm p,\varpi_l} \bm\psi^\dag(\bm p,\varpi_l) \,
  [i\varpi_l\hat\tau_0 - \xi(\bm p)\,\hat\tau_3
  \nonumber \\
  & - p_x\Delta_x\hat\tau_1 + p_y\Delta_y\hat\tau_2]
  \, \bm\psi(\bm p,\varpi_l).
  \end{align}
Here $\hat\tau_{1,2,3,0}$ are the Pauli matrices and the unity
matrix acting on the spinor $\bm\psi$. We set the Planck and
Boltzmann constants to unity: $\hbar=1$ and $k_B=1$.  The function
$\xi(\bm p)$ represents the electron energy dispersion counted from
the chemical potential $\mu$.  Interaction of electrons with the
electromagnetic field is described by the action $S_{\rm em}$
  \begin{align} \label{S_em}
  S_{\rm em} = -e\sum_{\bm p,\varpi_l} \bm v(\bm p) \!\cdot\! \bm A(i\omega_n)\,
  \bm\psi^\dag(\bm p,\varpi_l+\omega_n) \hat{\tau}_0 \bm\psi(\bm p,\varpi_l).
  \end{align}
Here $e$ is the electron charge, $\bm v(\bm p)=\partial\xi(\bm
p)/\partial\bm p$ is the electron velocity, $\omega_n$ is the
bosonic Matsubara frequency of the electromagnetic field, and the
wave vector $\bm q$ of the electromagnetic field is set to zero.

$\rm Sr_2RuO_4$ is a Q2D metal consisting of weakly coupled
conducting layers. Its electron dispersion $\xi(\bm p)$ can be
written as
  \begin{equation} \label{xi_p}
  \xi(\bm p) = \varepsilon_\|(p_x,p_y) - 2t_\perp\cos(p_z d) - \mu.
  \end{equation}
The first term in Eq.~\eqref{xi_p} represents the in-plane electron
dispersion, and the second term the tight-binding out-of-plane
dispersion with the tunneling amplitude $t_\perp$ and the interlayer
distance $d$.  Generally, the Fermi surface of $\rm Sr_2RuO_4$ is
rather complicated and consists of three sheets
\cite{Mackenzie_RevModPhys'03}.  To simplify presentation, we
consider only one sheet and assume that the in-plane dispersion is
isotropic, i.e.,\ $\varepsilon_\|$ depends only on
$p=\sqrt{p_x^2+p_y^2}$, which is a good approximation for the
$\gamma$ sheet \cite{Mackenzie_RevModPhys'03}.

We assume that the interlayer tunneling is weak $t_\perp\ll\mu $, so
the Fermi surface is a slightly corrugated cylinder extended along
the $p_z$ direction with the average radius $p_F$ in the
plane~\cite{maeno_PhysToday'01}.  For a given value of the
out-of-plane momentum $p_z$, Eq.~\eqref{xi_p} shows that the system
can be treated as a 2D metal with the effective chemical potential
$\tilde\mu(p_z)=\mu+2t_\perp\cos(p_z d)$ weakly dependent on $p_z$.
Such an effective 2D description is possible, because the pairing
potential $\Delta(\bm p)$ in Eq.\ (\ref{Delta_xy}) does not depend
on $p_z$, and the vector potential $\bm A$ in Eq.\ (\ref{S_em}) is
polarized in the $(x,y)$ plane.  So, in the rest of the paper, we
calculate the anomalous Hall conductivity $\sigma_{xy}$ per one
layer for a purely 2D electron system with the dispersion $\xi(\bm
p)=\varepsilon_\|(p)-\mu$.  A generalization to the Q2D case
involves trivial additional averaging over $p_z$ within each fermion
loop of the Feynman diagrams.  The bulk Hall conductivity is
obtained by dividing the 2D result for $\sigma_{xy}$ by the
interlayer distance $d$.

\subsection{Anomalous Hall conductivity of a chiral superconductor in the clean limit}  \label{Sec:Clean}

\begin{figure} \centering
\includegraphics[width=0.6\linewidth]{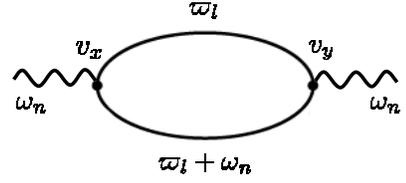}
  \caption{The Feynman diagram representing the anomalous Hall conductivity of a clean chiral $p$-wave superconductor. The wavy lines represent an external electromagnetic field, and the
solid lines are the electron Green's functions in the Nambu
representation.} \label{fig:no_line}
\end{figure}

The electromagnetic response function $Q_{xy}$ \eqref{Q} for the
model (\ref{S_el}) and (\ref{S_em}) is given by the Feynman diagram
in Fig.~\ref{fig:no_line}.  The analytic expression for this diagram
is
  \begin{align} \label{eq:response_clean}
  Q_{xy}(\omega_n)& = e^2 \mathrm{Tr} \left[ v_x(\bm p) \,
  \hat G(\varpi_l,\bm p) \, v_y(\bm p) \,
  \hat G(\varpi_l+\omega_n,\bm p) \right],
  \end{align}
where $\mathrm{Tr}$ denotes the trace over the Nambu space, as well
as the sum over the internal momenta $\bm p$ and fermionic
frequencies $\varpi_l$.  The Green's function for the chiral
$p$-wave superconductor in the Nambu representation is obtained by
inverting the kernel in Eq.~(\ref{S_el})
  \begin{align} \label{eq:Green's}
  \hat G(\varpi_l,\bm p)=-\frac{i\varpi_l\hat\tau_0 + \xi_{\bm p}\hat\tau_3
  + p_x\Delta_x\hat\tau_1 - p_y\Delta_y\hat\tau_2}
  {\varpi_l^2+\xi_{\bm p}^2+p_x^2\Delta_x^2+p_y^2\Delta_y^2}.
\end{align}
Here we write the argument $\bm p$ of the dispersion $\xi(\bm p)$ as the subscript in order to shorten the mathematical equation.

Using Eq.~(\ref{eq:Green's}) for calculation of the trace over the Pauli matrices $\hat\tau$ in Eq.~(\ref{eq:response_clean}), we find
  \begin{align} \label{eq:response_clean2}
  Q_{xy} = 2e^2 T \! \int \! \frac{d^2\bm p}{(2\pi)^2}
  v_x v_y \sum_{\varpi_l}
  \frac{i\varpi_l(i\varpi_l\!+\!i\omega_n)+\xi_{\bm p}^2+\Delta_0^2}
  {[(\varpi_l\!+\!\omega_n)^2\!+\!E_{\bm p}^2][\varpi_l^2\!+\!E_{\bm p}^2]},
  \end{align}
where $E_{\bm p}=\sqrt{\xi_{\bm p}^2+\Delta_0^2}$ is the
quasiparticle energy.  It is clear that the response function
$Q_{xy}$ in Eq.~(\ref{eq:response_clean2}) vanishes upon integration
over the orientation of $\bm p$, because $v_x$ is an odd function of
$p_x$, and $v_y$ is an odd function of $p_y$.  Even if an
anisotropic model with $\Delta_x\neq\Delta_y$ is
considered~\cite{kim_prl'08}, still a similar calculation gives a
symmetric tensor $Q_{xy}=Q_{yx}$, which does not represent the Hall
conductivity~\cite{Joynt_PRB'91} and is not relevant for the
experiment~\cite{xia_prl'06}.

\subsection{Disorder-induced anomalous Hall conductivity}
\label{Sec:disorder}

As discussed in Sec.~\ref{Sec:Impurities}, in order to obtain a
non-zero anomalous Hall conductivity for a chiral superconductor, it
is necessary to include the effect of disorder. Thus, we add the
impurity scattering term to the action of the system
  \begin{align} \label{S_imp}
  S_{\rm imp} = -\!\sum_{\bm q,\bm p,\varpi_l} V_{\rm imp}(\bm q)\,
  \bm\psi^\dag(\bm p+\bm q,\varpi_l) \hat{\tau}_3 \bm\psi(\bm p,\varpi_l),
  \end{align}
where $V_{\rm imp}(\bm q)$ is the impurity potential written in the
momentum representation. We assume that the dominant scattering
mechanism comes from the short-range disorder.  The first and the
second moments of the probability distribution function of the
impurity potential $V_{\rm imp}(\bm q)$ are
  \begin{align} \label{eq:sec_cumulant}
  \langle V_{\rm imp}(\bm q) \rangle & = 0,
  \nonumber \\
  \langle V_{\rm imp}(\bm q) \, V_{\rm imp }(\bm q') \rangle
  & = n_i u_0^2 \, \delta(\bm q + \bm q').
  \end{align}
Here the averaging is performed over different realizations of the
disordered potential, and $n_i$ and $u_0$ are the 2D concentration
of impurities and the strength of the disorder potential,
respectively.

It is well known that impurity scattering suppresses unconventional
superconducting pairing when the parameter $\Delta_0\tau$ (where
$\tau$ is the quasiparticle scattering time) becomes of the order of
unity, because the Anderson theorem does not apply to non-$s$-wave
pairing~\cite{balatsky_RevModPhys'06}. Suppression of
superconductivity by disorder has been experimentally observed in
$\rm Sr_2RuO_4$~\cite{Mackenzie_PRL'98} and is one of the arguments
in favor of the unconventional pairing symmetry. Therefore, we study
the case where the concentration of impurities is very low, so that
$\Delta_0\tau\gg1$, and the superconducting pairing is not
significantly affected by impurities.  In fact, $\rm Sr_2RuO_4$ is a
very clean stoichiometric material, where impurities are not
introduced intentionally, and physical origin of residual disorder
is not clear.  Nevertheless, the presence of a small, but non-zero
concentration of impurities can still induce a non-zero anomalous
Hall effect.  For such a low concentration of impurities, the
anomalous Hall response can be studied perturbatively in the
strength of the disorder potential and would be dominated by the the
lowest-order non-vanishing diagrams.

\begin{figure} \centering
\includegraphics[width=1.0\linewidth]{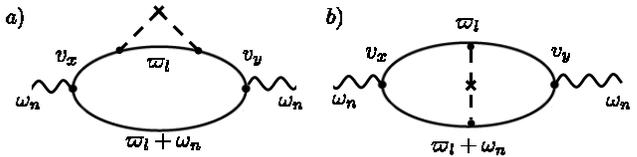}
  \caption{The lowest-order diagrams in impurity scattering for the anomalous Hall conductivity.  The dashed lines represent scattering potential from the same impurity, shown by the cross.} \label{fig:single_line}
\end{figure}

The lowest-order contributions to the anomalous Hall conductivity
appear in the second order in $V_{\rm imp }(\bm q)$ as shown in
Fig.~\ref{fig:single_line}.  The diagram a) represents a self-energy
correction to the Green's function of the electrons due to disorder.
In $p$-wave superconductors, the  momentum average of the gap is
zero: $\sum_{\bm p}\Delta(\bm p)=0$.  Thus, the self-energy
corrections modify only  the $\tau_0$ and $\tau_3$ components of the
Green's function (\ref{eq:Green's}).  However, such a modification
does not change the Nambu structure of the current-current
correlation function (\ref{eq:response_clean}).  Therefore,
similarly to the clean case, the diagram a) gives a vanishing
contribution to the anomalous Hall conductivity.

On the other hand, the diagram b) in Fig.~\ref{fig:single_line} has
a non-trivial structure and deserves a more detailed discussion.
The analytical expression for the diagram b) is
  \begin{align} \label{Q1b}
  Q_{xy}^{(2b)}(\omega_n) = n_i u_0^2\, \mathrm{Tr}
  \left[ \hat\Lambda_x \hat\tau_3 \hat\Lambda_y \hat\tau_3 \right],
  \end{align}
where $\mathrm{Tr}$ is taken over the Pauli matrices and the
internal frequency $\varpi_l$. The effective vertices
$\hat\Lambda_x$ and $\hat\Lambda_y$ are
  \begin{align} \label{Lambda_x}
  \hat\Lambda_x &= -e \sum_{\bm p_1} \hat G(\varpi_l+\omega_n,\bm p_1) \,
  v_{x}(\bm p_1) \, \hat G(\varpi_l,\bm p_1)
  \\
  &= -\frac{e p_F^2}{2} \Delta_x \hat\tau_1 \int_{-\omega_D}^{\omega_D}
  \frac{d\xi}{2\pi} \frac{i(2\varpi_l+\omega_n)}
  {[(\varpi_l+\omega_n)^2+E_{\bm p}^2][\varpi_l^2+E_{\bm p}^2]}
  \nonumber \\
  &\approx -\frac{i e p_F^2 \Delta_x}{4\omega_n} \hat\tau_1
  \left(\frac{1}{\sqrt{\varpi_l^2+\Delta_0^2}}
  -\frac{1}{\sqrt{(\varpi_l+\omega_n)^2+\Delta_0^2}}\right)
  \nonumber
  \end{align}
and
  \begin{align} \label{Lambda_y}
  \hat\Lambda_y &= -e \sum_{\bm p_2} \hat G(\varpi_l,\bm p_2) \,
  v_{y}(\bm p_2) \, \hat G(\varpi_l+\omega_n,\bm p_2)
  \\
  &= i \frac{e p_F^2}{2} \Delta_y \hat\tau_2 \int_{-\omega_D}^{\omega_D}
  \frac{d\xi}{2\pi} \frac{(2\varpi_l+\omega_n)}
  {[(\varpi_l+\omega_n)^2+E_{\bm p}^2][\varpi_l^2+E_{\bm p}^2]}
  \nonumber\\
  &\approx \frac{i e p_F^2 \Delta_y}{4\omega_n} \hat\tau_2
  \left(\frac{1}{\sqrt{\varpi_l^2+\Delta_0^2}}
  -\frac{1}{\sqrt{(\varpi_l+\omega_n)^2+\Delta_0^2}}\right).
  \nonumber
  \end{align}
In going from the first to the second lines in Eqs.\
(\ref{Lambda_x}) and (\ref{Lambda_y}), we integrated over the
angular orientation of the electron momentum $\bm p$.  For the
integration in the radial direction, we make the linearized
approximation for the 2D electron dispersion
  \begin{align} \label{linearized}
  \xi(\bm p) \approx v_F (p-p_F),
  \end{align}
where $v_F$ is the Fermi velocity.  This approximation is justified,
because the integrals converge in the vicinity of the Fermi surface.
Then, the integration over $dp$ can be replaced by integration over
$d\xi=v_Fdp$.  The limits of integration over $\xi$ are given by the
BCS cutoff energy $\omega_D$, which is determined, presumably, by
the energy scale of ferromagnetic fluctuations inducing the $p$-wave
superconductivity in $\rm Sr_2RuO_4$.  Throughout the paper, we
assume that $\omega_D\gg\omega,\Delta_0$ and neglect corrections to
Eqs.\ (\ref{Lambda_x}) and (\ref{Lambda_y}) of the order
$O(\omega_D^{-3})$.

Substituting Eqs.\ (\ref{Lambda_x}) and (\ref{Lambda_y}) into
Eq.~(\ref{Q1b}), we find that $Q_{xy}^{(2b)}=0$ because $\mbox{Tr}\{
\hat\tau_1 \hat\tau_3 \hat\tau_2 \hat\tau_3 \}=0$.  So, there is no
contribution to the Hall conductivity to the second order in $V_{\rm
imp}$.  Thus, we have to consider the higher-order diagrams in order
to obtain a non-zero Hall conductivity.

\subsection{Anomalous Hall conductivity due to the higher-order diagrams in disorder potential}

\begin{figure} \centering
\includegraphics[width=0.99\linewidth]{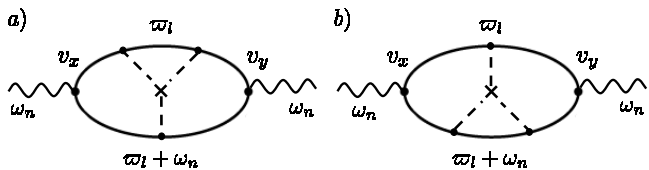}
  \caption{The skew-scattering diagrams contributing to the anomalous Hall conductivity.  The three connected dashed lines represent the third moment \eqref{3rd-moment} of the impurity potential.} \label{fig:skew_diagrams}
\end{figure}

In the rest of the paper, we study the higher-order contributions in
disorder potential.  This requires to specify the assumptions about
the probability distribution of the disorder potential $V_{\rm
imp}$.  Below, we consider two models with the Gaussian and
non-Gaussian distributions of the disorder potential. In the case of
the Gaussian distribution of disorder, it is enough to specify the
second-order cumulant given in Eq.~(\ref{eq:sec_cumulant}).  All
other higher-order moments are either zero (odd in $V_{\rm imp }$)
or can be expressed in terms of the second-order cumulant (even in
$V_{\rm imp }$).  In the case of a non-Gaussian distribution of
disorder, the odd cumulants may be non-zero.  In particular, the
third moment (skewness) within a non-Gaussian model is
  \begin{align} \label{3rd-moment}
  \langle V_{\rm imp }(\bm q)\,V_{\rm imp }(\bm q')\,V_{\rm imp }(\bm q'')
  \rangle = \kappa_3 n_i u_0^3\, \delta(\bm q+\bm q'+\bm q'').
  \end{align}
Here $\kappa_3$ is a dimensionless parameter characterizing the
skewness, which varies from 0 to 1 depending on the deviation of the
actual distribution function from the Gaussian.  For example,
$\kappa_3=1$ corresponds to randomly distributed impurities
generating short-range, delta-function potentials of the equal
strength $u_0$.  The contributions to the Hall conductivity to the
first order in the third moment (\ref{3rd-moment}) are shown in
Fig.~\ref{fig:skew_diagrams}.

The skew-scattering diagrams, similar to those shown in
Fig.~\ref{fig:skew_diagrams}, were studied for the anomalous Hall
effect in
ferromagnets~\cite{smit_physica'55,nozieres_JdP'73,sinitsyn_prb'07}
and for the extrinsic spin Hall
effect~\cite{dyakonov_JETP'71,Engel_PRL'05,tse_PRL'06} in
semiconductors.  In the case of a chiral $p$-wave superconductor,
the anomalous Hall conductivity due to the skew-scattering diagrams
of Fig.~\ref{fig:skew_diagrams} was first calculated by
Goryo~\cite{goryo_prb'08} for high frequencies $\omega\gg\Delta_0$
and $T$ close to $T_c$.  In Sec.\ \ref{sec:skew}, we calculate the
real and imaginary parts of $\sigma_{xy}(\omega)$ originating from
the skew-scattering diagrams shown in Fig.~\ref{fig:skew_diagrams}
for the full range of frequencies $\omega$ and for any temperature
$T$.

\begin{figure} \centering
\includegraphics[width=0.99\linewidth]{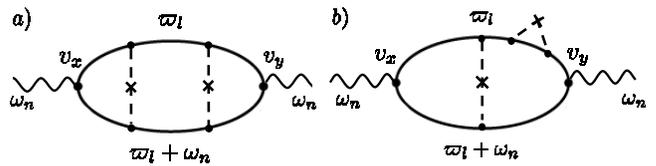}
  \caption{The lowest-order diagrams contributing to the Hall conductivity within the Gaussian model of disorder with the particle-hole asymmetry.}
  \label{fig:2nd_order_line}
\end{figure}

For the Gaussian model of disorder, the higher-order diagrams
contributing to the Hall conductivity are shown in
Fig.~\ref{fig:2nd_order_line}.  We calculate the real and imaginary
parts of $\sigma_{xy}(\omega)$ originating from these diagrams in
Sec.~\ref{sec:p-hasym}.  We show that these diagrams give a non-zero
contribution to the Hall conductivity only when we take into account
the particle-hole asymmetry, which is discussed in more detail in
the next Subsection.

\subsection{Sign of the anomalous Hall conductivity}\label{sec:sign}

In this Subsection, we use the symmetries of the problem to discuss
what factors determine the sign of the anomalous Hall conductivity.

First, let us discuss the time-reversal operation.  The Hall
conductivity changes sign upon the time reversal:
$\sigma_{xy}\to-\sigma_{xy}$.  On the other hand, the time-reversal
operation also results in complex conjugation of the Hamiltonian.
For a chiral $p$-wave superconductor, it means that $\Delta(\bm
p)\to\Delta^*(\bm p)$, so the pairing potential (\ref{Delta_xy})
changes its chirality.  Thus, we conclude that
  \begin{align} \label{TRS}
  \sigma_{xy}(\Delta)=-\sigma_{xy}(\Delta^*),
  \end{align}
i.e., changing the sign of chirality of the superconducting order
parameter changes the sign of the anomalous Hall conductivity, so
$\sigma_{xy}\propto s_{xy}$, see Eq.~(\ref{s_xy}).

When we employ the commonly used linearized approximation
(\ref{linearized}) for the electron dispersion, the problem acquires
an additional symmetry upon exchange of the electron and hole
operators, $\psi(\bm p)\to\psi^\dag(\bm p)$ and $\psi^\dag(\bm
p)\to\psi(\bm p)$.  This operation preserves the fermion
anticommutation relations, so it is a canonical transformation. Upon
the particle-hole transformation and subsequent commutation of the
$\psi$ and $\psi^\dag$ operators, the Hamiltonian of the kinetic
energy of electrons $\sum_{\bm p}\xi_{\bm p}\psi^\dag(\bm p)\psi(\bm
p)$ transforms to the original form (up to an additive constant),
but with the sign change of the energy dispersion $\xi_{\bm
p}\to-\xi_{\bm p}$.  If, in addition, we use the approximation
(\ref{linearized}) and change the radial momentum variables
$p\to2p_F-p$ in the sum, the kinetic energy returns to the original
form, i.e., becomes invariant under the particle-hole
transformation.

Now, let us examine how the other terms in the Hamiltonian transform
upon the particle-hole transformation
$\psi\leftrightarrow\psi^\dag$.  For the superconducting pairing
potential, we find that $\Delta\to\Delta^*$, so the chirality
changes to the opposite.  For the impurity potential (\ref{S_imp}),
we find that $V_{\rm imp}\to-V_{\rm imp}$.  On the other hand, the
observable Hall conductivity $\sigma_{xy}$ should not depend on the
choice of operators $\psi$ and $\psi^\dag$, which are integrated
out.  So, we obtain
  \begin{align} \label{e-h-symm}
  \sigma_{xy}(\Delta,V_{\rm imp})=\sigma_{xy}(\Delta^*,-V_{\rm imp})
  \end{align}
and, combining with Eq.~(\ref{TRS}),
  \begin{align} \label{V=>-V}
  \sigma_{xy}(\Delta,V_{\rm imp})=-\sigma_{xy}(\Delta,-V_{\rm imp}).
  \end{align}
Eqs.~(\ref{e-h-symm}) and (\ref{V=>-V}) are valid only when the
approximation (\ref{linearized}) is employed, and the electron
kinetic energy has the particle-hole symmetry.

Eq.~(\ref{V=>-V}) shows that the sign of the Hall conductivity must
change with the sign change of the impurity potential.  This is
indeed the case for the skew-scattering diagrams shown in
Fig.~\ref{fig:skew_diagrams}, where $\sigma_{xy}\propto V_{\rm
imp}^3$.  Thus, we conclude that the skew-scattering diagrams can be
calculated using the linearized approximation (\ref{linearized}).
The sign of the anomalous Hall conductivity given by these diagrams
changes when repulsive impurities are replaced by attractive ones.
In that sense, the sign of the anomalous Hall effect produced by
skew scattering is not a property of the $\rm Sr_2RuO_4$ material as
such, but is a property of the impurities in this material.  This
conclusion is contrast to the sign of the conventional Hall effect,
which is determined by the sign of carriers, electrons or holes, but
not by the sign of the impurity potential.

On the other hand, the diagrams for the Gaussian model of disorder,
shown in Fig.~\ref{fig:2nd_order_line}, give $\sigma_{xy}\propto
V_{\rm imp}^4$, which is incompatible with Eq.~(\ref{V=>-V}).  It
means that these diagrams must vanish when calculated using the
linearized approximation (\ref{linearized}). In order to obtain a
non-zero result for these diagrams, we must go beyond the linearized
approximation (\ref{linearized}) and take into account the curvature
of the electron dispersion $\xi(p)$, i.e.,\ to take into account the
electron-hole asymmetry.  This is similar to the conventional Hall
effect, whose sign is determined by the particle-hole asymmetry.  In
this sense, the anomalous Hall effect given by the diagrams in
Fig.~\ref{fig:2nd_order_line} reflects the properties of the
material, because its sign is determined by asymmetry of the
electron spectrum and does not depend on the sign of the impurity
potential.

It should be emphasized, however, that the magnitude of the
anomalous Hall conductivity, given by the diagrams in
Figs.~\ref{fig:skew_diagrams} and \ref{fig:2nd_order_line}, is still
proportional to some power of the strength of the impurity
scattering.  In this sense, the magnitudes of the measured anomalous
Hall conductivity and the Kerr angle do not characterize the $\rm
Sr_2RuO_4$ material as such, but characterize the degree of disorder
in the samples of this material.  This conclusion can be verified
experimentally by intentionally introducing mild dozes of disorder
into $\rm Sr_2RuO_4$, such that superconductivity is not
significantly suppressed.  The theory predicts that the observed
Kerr angle should increase substantially after introduction of
disorder.

Concluding this Section, we remark that the diagrams shown in
Fig.~\ref{fig:single_line} still vanish, even if we take into
account the particle-hole asymmetry.  Their vanishing is a
consequence of angular integration over electron momentum.  We also
mention that the effects of particle-hole asymmetry are important to
explain certain properties of the superfluid
$^3$He~\cite{Leggett_RevModPhys'75}, and, thus, may be relevant for
$\rm Sr_2RuO_4$ as well.  By taking into account the particle-hole
asymmetry, Yip and Sauls~\cite{Yip_JLTP'92} have shown that a chiral
$p$-wave superconductor exhibits circular dichroism and
birefringence, which arise from the collective modes of the order
parameter.  However, the effect is too small to explain the
experiment~\cite{xia_prl'06}.  The particle-hole asymmetry was also
discussed in relation with the sign change of the Hall effect at
$T_c$ in the high-$T_c$ superconductors attributed to vortex
motion~\cite{Kopnin_JETP'92,Feigelman_JETPLett'95,Otterlo_PRL'95,Volovik_CJP'96}.

\section{Skew scattering in the non-Gaussian model of disorder}
\label{sec:skew}

In this Section, we study the non-Gaussian model of disorder. Within
this model, the non-zero contributions to the anomalous Hall
conductivity $\sigma_{xy}$ are given by the diagrams in
Fig.~\ref{fig:skew_diagrams}.  These diagrams differ from the
diagram in Fig.~\ref{fig:single_line} by the presence of an
additional fermion loop.  The analytical expressions for the
diagrams a) and b) in Fig.~\ref{fig:skew_diagrams} can be written as
  \begin{align}
  Q_{xy}^{(3a)}(\omega_n) &= \kappa_3 n_i u_0^3\mathrm{Tr}
  \left[ \hat\Lambda_x  \hat\tau_3 \hat G(\varpi_l,\bm p_3) \hat\tau_3
  \hat\Lambda_y \hat\tau_3 \right] ,
  \label{trip_a2} \\
  Q_{xy}^{(3b)}(\omega_n) &= \kappa_3 n_i u_0^3\mathrm{Tr}
  \left[ \hat\Lambda_x \hat\tau_3 \hat\Lambda_y \hat\tau_3
  \hat G(\varpi_l\!+\!\omega_n,\bm p_3) \hat\tau_3 \right]\!,
  \label{trip_b2}
  \end{align}
where the effective vertices $\hat\Lambda_x$ and $\hat\Lambda_y$ are given by Eqs.\ (\ref{Lambda_x}) and (\ref{Lambda_y}).  The trace is taken over the Pauli matrices, the frequency $\varpi_l$, and the momentum $\bm p_3$ in the additional fermion loop.  The integration over $\bm p_3$ yields
  \begin{align} \label{loop}
  \sum_{\bm p_3} \hat G(\varpi_l,\bm p_3) \!=\! N(0) \!\!\!
  \int\limits_{-\omega_D}^{\omega_D} \!\! d\xi
  \frac{- i \varpi_l \hat\tau_0}{\varpi_l^2+\xi^2+\Delta_0^2}
  \approx \frac{- i \varpi_l \hat\tau_0 N(0)\pi}{\sqrt{\varpi_l^2+\Delta_0^2}}.
  \end{align}
Here we used the approximation \eqref{linearized} and introduced the 2D energy density of states $N(0)$ at the Fermi level
  \begin{align} \label{N(0)}
  N(0) = \frac{1}{V}\sum_{p}\delta[  \xi(\bm p) ]\approx \frac{p_F}{2\pi v_F},
  \end{align}
where $V$ is the volume of the system. For a model with the parabolic dispersion $\varepsilon_\|=p^2/2m_e$ and an effective electron mass $m_e$, Eq.~\eqref{N(0)} gives $N(0)=m_e/2\pi$, but we will write our results for a general dispersion.

Substituting Eq.~\eqref{loop} into Eqs.~(\ref{trip_a2}) and (\ref{trip_b2}) and evaluating the Nambu traces ($Q_{xy}^{(3a)} \propto {\rm Tr}\{\hat \tau_1 \hat \tau_2 \hat \tau_3\}=2i$ and  $Q_{xy}^{(3b)} \propto {\rm Tr}\{\hat \tau_1 \hat \tau_3 \hat \tau_2\}=-2i$), we find the following result for the full response function $Q_{xy}^{(3)}=Q_{xy}^{(3a)}+Q_{xy}^{(3b)}$ \cite{footnote3}:
  \begin{align} \label{Q3}
  Q_{xy}^{(3)}(\omega_n) = \kappa_3 n_i u_0^3
  \frac{e^2 \pi N(0) p_F^4 \Delta_x \Delta_y}{8\omega_n^2 \Delta_0} H(\omega_n).
  \end{align}
Here the dimensionless function $H(\omega_n)$ is
  \begin{align}
  & H(\omega_n) = \Delta_0 T \sum_{l}\! \left(
  \!\frac{1}{\sqrt{\varpi_l^2+\Delta_0^2}}\!
  -\!\frac{1}{\sqrt{(\varpi_l\!+\!\omega_n)^2\!+\!\Delta_0^2}}\!\right)^2
  \nonumber\\
  &\times \left( \frac{\varpi_l}{\sqrt{\varpi_l^2\!+\!\Delta_0^2}}
  -\frac{\varpi_l +\omega_n}{\sqrt{(\varpi_l+\omega_n)^2+\Delta_0^2}} \right).
  \label{trip}
  \end{align}
The Matsubara sum in Eq.~(\ref{trip}) can be evaluated using a contour of integration in the complex plane enclosing the points of non-analytic behavior, as shown in Appendix~\ref{App:eval_I}.  The result is
  \begin{align} \label{eq:Ifinal}
  & \! H(\omega_n)\! =
  \!\frac{\tanh\!\left(\!\frac{\Delta_0}{2T}\!\right)}{2}\!\!
  \left[
  \!\frac{\omega_n\!+\!3i\Delta_0}{\sqrt{\omega_n(\omega_n\!+\!2i\Delta_0)}}
  \!+\!
  \frac{\omega_n\!-\!3i\Delta_0}{\sqrt{\omega_n(\omega_n\!-\!2i\Delta_0)}}
  \!\right]
  \\
  & \!+\!\frac{\Delta_0}{\pi}\int\limits^{\infty}_{\Delta_0}
  \frac{\tanh\!\left(\!\frac{x}{2T}\!\right)\! dx}
  {\sqrt{x^2\!-\!\Delta_0^2}} \!
  \left[ \frac{3ix\!-\!2\omega_n}{\Delta_0^2\!+\!(ix\!-\!\omega_n)^2} \!-\!
  \frac{3ix\!+\!2\omega_n}{\Delta_0^2\!+\!(ix\!+\!\omega_n)^2} \right] \! .
  \nonumber
  \end{align}
Then, we perform the analytical continuation in Eq.~(\ref{eq:Ifinal})
  \begin{align} \label{Omega+}
  i\omega_n \to \omega^+ = \omega+i\delta,
  \end{align}
where $\omega$ is the real physical frequency, and $\delta$ is an infinitesimal positive number.  Thus we obtain the retarded finite-temperature response function $Q_{xy}^{(3)}(\omega)$ and, via Eq.~(\ref{eq:sigma-q}), the anomalous Hall conductivity $\sigma^{(3)}_{xy}(\omega)$ per layer
  \begin{align} \label{eq:sigma3}
  \sigma^{(3)}_{xy}(\omega) &= s_{xy} \frac{e^2}{\hbar} \,
  \frac{W\Delta_0}{\omega^3} \,
  H\left(\frac{\omega^+}{\Delta_0}\right),
  \\
  W &= -\kappa_3 n_i u_0^3 \, \frac{\pi N(0) p_F^2}{8 \hbar^2}.
  \label{W}
  \end{align}
Here the sign function $s_{xy}$ is given by Eq.~(\ref{s_xy}).  We restore the factor $\hbar$ in Eqs.~\eqref{eq:sigma3} and \eqref{W}, but assume that the frequency $\omega$ is measured in the energy units.  We also introduce the parameter $W$ with the dimensionality of (energy)$^2$, which characterizes the strength and skewness of the non-Gaussian disorder potential.  The parameter $W$ can be positive or negative depending on the sign of the impurity potential $u_0$ and will be estimated in Sec.~\ref{sec:PKE}.  The dimensionless function $H(x)$ describes frequency and temperature dependence of the chiral response
  \begin{align}
  & H(x) = \int\limits^{\infty}_1 dy
  \frac{\tanh\left(\!\frac{y\Delta_0}{2T}\!\right)}{\pi\sqrt{y^2-1}}
  \left[\frac{3y+2x}{1\!-\!(y\!+\!x)^2} - \frac{3y-2x}{1\!-\!(y\!-\!x)^2}\right]
  \nonumber\\
  & - \frac12 \tanh\left(\!\frac{\Delta_0}{2T}\!\right)
  \left[\frac{x-3}{\sqrt{x(2\!-\!x)}} + \frac{x+3}{\sqrt{-x(x\!+\!2)}}\right].
  \label{F}
  \end{align}
Generally, the function $H(x)$ takes complex values when the variable $x=\omega^+/\Delta_0$ changes from $-\infty$ to $\infty$.   Using  Eqs.~(\ref{F}) and (\ref{eq:sigma3}), it is easy to check that $\sigma_{xy}(\omega)$ satisfies the causality requirement: $\sigma_{xy}(\omega)=\sigma_{xy}^*(-\omega)$~\cite{Landau_book8}.  Eqs.\ (\ref{eq:sigma3}) and (\ref{F}) give a complete answer for the real and imaginary parts of the anomalous Hall conductivity $\sigma^{(3)}_{xy}(\omega)$ for arbitrary frequency and temperature.

Now we focus on the limit $T=0$, where the function $H(x)$ can be calculated analytically by setting the $\tanh$ function to unity in Eq.~(\ref{F})
  \begin{align} \label{eq:Fx}
  H(x) = & - \frac{(x-3)\arccos(1-x)}{\pi\sqrt{x(2-x)}}
  - \frac{(x+3)\arccos(1+x)}{\pi\sqrt{-x(2+x)}}.
  \end{align}
Taking into account that the function $\arccos(x)$ of a real variable $x$ has a real value for $|x|<1$ and an imaginary value for $|x|>1$, we observe that the second term in Eq.~(\ref{eq:Fx}) is always real for $x>0$, whereas the first term is real for $0<x<2$ and imaginary for $x>2$.  The appearance of the imaginary part for $x>2$ represents the threshold of absorption across the superconducting gap for photons with the energies $\omega>2\Delta_0$.  The real and imaginary parts of $H(x)$ exhibit a square-root singularity at $x=2$: $H(x)\sim 1/\sqrt{2-x}$.  The other asymptotes are
  \begin{align} \label{eq:asymptotes}
  H(x)= \left\{
  \begin{array}{ll}
  \displaystyle \frac{8}{105\pi} \, x^3, & x \to 0
  \\ \\
  \displaystyle
  -i - \frac{4\ln x}{\pi x}, & x \to \infty.
  \end{array}  \right.
  \end{align}

\begin{figure} \centering
\includegraphics[width=0.99\linewidth]{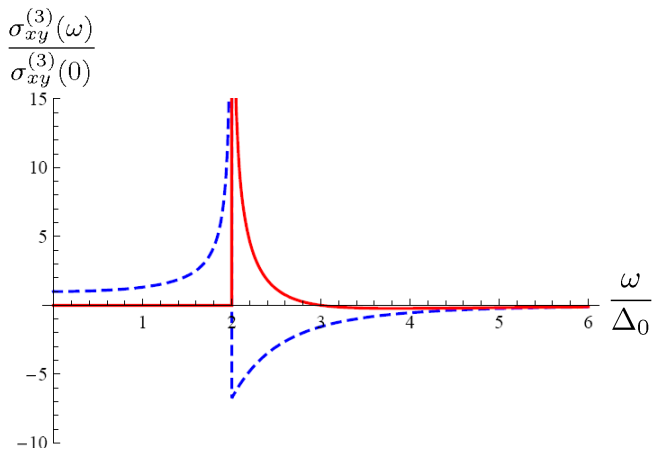}
  \caption{(Color online) Frequency dependence of the real (dashed line) and imaginary (solid line) parts of the anomalous Hall conductivity $\sigma^{(3)}_{xy}(\omega)$ at $T=0$ given by Eqs.~\eqref{eq:sigma3} and \eqref{eq:Fx}. } \label{fig:FOm}
\end{figure}

Frequency dependence of the real and imaginary parts of $\sigma_{xy}^{(3)}(\omega)$ at $T=0$, obtained from Eqs.\ (\ref{eq:sigma3}) and (\ref{eq:Fx}), is plotted in Fig.~\ref{fig:FOm}.  We observe that the imaginary part of $\sigma_{xy}$ appears only for $\omega>2\Delta_0$, above the threshold of photon absorption.  In the high-frequency limit $\omega\gg\Delta_0$, $\sigma_{xy}^{(3)}(\omega)$ is given by Eqs.\ (\ref{eq:sigma3}) and \eqref{eq:asymptotes}
  \begin{align} \label{Hall_skew_high}
  \sigma^{(3)}_{xy}(\omega) = - s_{xy} \, \frac{e^2}{\hbar} \,
  \frac{W\Delta_0}{\omega^3} \, \left[i + \frac{4}{\pi}
  \frac{\Delta_0}{\omega} \ln\left(\frac{\omega}{\Delta_0} \right) \right],
  \end{align}
where $\Delta_0$ is the superconducting gap at $T=0$.  In the dc limit $\omega=0$, $\sigma_{xy}^{(3)}$ has the finite real value
  \begin{align} \label{Hall_skew_dc}
  \sigma^{(3)}_{xy}(\omega=0)= s_{xy} \, \frac{8}{105\pi} \, \frac{e^2}{\hbar} \,
  \frac{W}{\Delta_0^2} .
  \end{align}

\begin{figure} \centering
\includegraphics[width=0.99\linewidth]{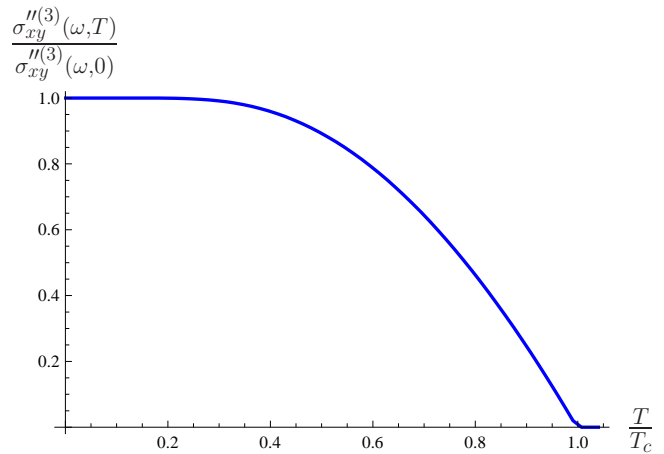}
  \caption{(Color online) Temperature dependence of the imaginary part of the anomalous Hall conductivity $\sigma''^{(3)}_{xy}(\omega,T)$ for $\omega\gg T_c$ given by Eq.~\eqref{eq:sigma3T} with $\Delta_0(T)$ from the BCS theory \cite{footnote:theta(T)}. } \label{fig:kerr-T}
\end{figure}

At a finite temperature $T\neq0$, $\Delta_0(T)$ in Eqs.\ (\ref{eq:sigma3}) and (\ref{F}) should be understood as the temperature-dependent superconducting energy gap obtained by solving the appropriate BCS equation for the chiral $p$-wave pairing~\cite{Leggett_RevModPhys'75}.  It can be easily verified that, at high frequency $\omega\gg T_c$, the imaginary part of $\sigma^{(3)}_{xy}(\omega)$ is much greater than the real part at any $T<T_c$, as in Eq.~\eqref{Hall_skew_high} at $T=0$.  Thus, we focus on temperature dependence of the imaginary part $\sigma''^{(3)}_{xy}(\omega,T)$.  Taking the limit $x\gg1$ in Eq.~(\ref{F}) and observing that the first, integral term gives a negligible contribution to the imaginary part, we find from Eq.~(\ref{eq:sigma3})
  \begin{align} \label{eq:sigma3T}
  \frac{\sigma''^{(3)}_{xy}(\omega,T)}{\sigma''^{(3)}_{xy}(\omega,0)}
  = \frac{\Delta_0(T)}{\Delta_0(0)} \,
  \tanh\left( \frac{\Delta_0(T)}{2T} \right),
  \quad \omega\gg T_c.
  \end{align}
The plot of Eq.~\eqref{eq:sigma3T} is shown in Fig.~\ref{fig:kerr-T} using $\Delta_0(T)$ calculated from the BCS equation \cite{book_abrikosov, footnote:theta(T)}.  For $T$ close to $T_c$, where $\Delta_0(T)\ll T$, the $\tanh$ function in Eq.~\eqref{eq:sigma3T} can be replaced by its argument, so $\sigma^{(3)}_{xy}$ becomes proportional to the square of the superconducting order parameter $\Delta_0^2(T)$.  In the Ginzburg-Landau theory, $\Delta_0^2(T)\propto(T_c-T)$, so Eq.~\eqref{eq:sigma3T} gives a linear temperature dependence for $\sigma^{(3)}_{xy}$ (and, thus, $\theta_K$) near $T_c$.  The factor $\Delta_0^2$ originates from the product $\Delta_x\Delta_y$ in Eq.~\eqref{Q3}, because both $\Delta_x$ and $\Delta_y$ are necessary for a non-zero anomalous Hall effect, as emphasized in Ref.~\cite{yakovenko_prl'07}.  However, $\sigma^{(3)}_{xy}$ at $T=0$ in Eq.~(\ref{Hall_skew_high}) is proportional to the first power of $\Delta_0$, somewhat reminiscent to what was proposed in Ref.~\cite{xia_prl'06}.  Curiously, the temperature dependence in Eq.~(\ref{eq:sigma3T}) is the same as for the critical current in Josephson junctions~\cite{Ambegaokar_PRL'63}.

For high frequency $\omega\gg T_c$ and $T$ close to $T_c$, we obtain the following expression for the real and imaginary parts of $\sigma^{(3)}_{xy}(\omega,T)$ from Eqs.~(\ref{eq:sigma3}) and (\ref{F})
  \begin{align} \label{T->Tc}
  \sigma^{(3)}_{xy}(\omega,T) = - s_{xy} \frac{e^2}{\hbar} \,
  \frac{W\Delta_0^2(T)}{\omega^3 \, T_c} \, \left[ \frac{i}{2}
  + \frac{4}{\pi}
  \frac{T_c}{\omega} \ln\left(\frac{\omega}{T_c} \right) \right].
  \end{align}
The imaginary part $\sigma^{\prime\prime}_{xy}\propto1/\omega^3$ in Eq.~(\ref{T->Tc}) looks similar to the real part of $\sigma_{xy}$ calculated by Goryo in Eq.~(5) of Ref.~\cite{goryo_prb'08} for the same model in the same asymptotic limit.  Given the causality requirement $\sigma_{xy}(\omega)=\sigma_{xy}^*(-\omega)$~\cite{Landau_book8}, it appears that the factor of $i=\sqrt{-1}$ was overlooked in the calculation of Ref.~\cite{goryo_prb'08}, so the real and imaginary parts of $\sigma_{xy}$ were interchanged.

\section{Model of the Gaussian disorder with particle-hole asymmetry}
\label{sec:p-hasym}

In this section, we calculate the anomalous Hall conductivity for the Gaussian  model of disorder by taking into account the effects of particle-hole asymmetry.

The lowest-order non-vanishing diagrams in this case are shown in Fig.~\ref{fig:2nd_order_line}.  Fig.~\ref{fig:2nd_order_line}b actually represents four similar diagrams, with self-energy corrections to the upper and lower Green's functions.  These diagrams can be considered as the lowest-order terms of a more general diagrammatic series shown in Fig.~\ref{fig:self_vertex}, which is often considered in calculations of transport properties of metals with disorder.  We find it more practical to calculate the general series of diagrams in Fig.~\ref{fig:self_vertex} and then take the limit of low concentration of impurities, which corresponds to the diagrams in Fig.~\ref{fig:2nd_order_line}.  In this way, we can be sure that all contributions of the same order are taken into account, and mutual signs of the diagrams are correct.

\subsection{Self-consistent non-crossing approximation}

\begin{figure} \centering
\includegraphics[width=0.9\linewidth]{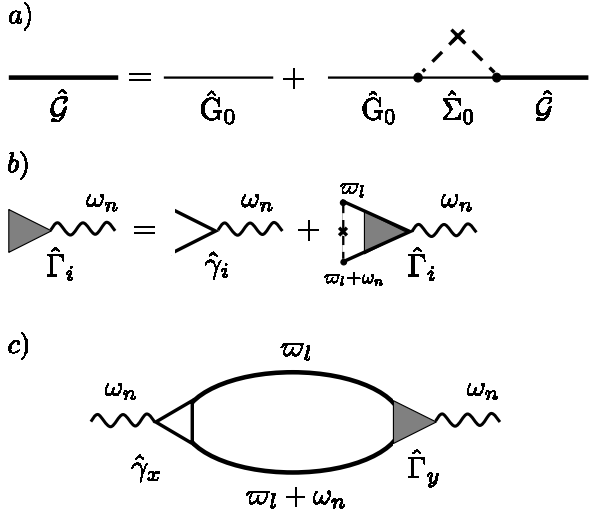}
  \caption{ Diagrams for the disorder-dressed Green's function (a), the vertex function (b), and the response function (c). } \label{fig:self_vertex}
\end{figure}

We begin with calculation of the Green's function $\hat{\cal G}_{\bm p}(\varpi_l)$ dressed due to impurity scattering.  (Starting from this Section, we write the momentum $\bm p$ in the argument of a Green's function as a subscript in order to shorten notation in long mathematical equations.)  The dressed Green's function is obtained from  Dyson's equation shown in Fig.~\ref{fig:self_vertex}a, where we make the standard non-crossing approximation, assuming that $E_F\tau\gg1$ and neglecting diagrams with intersecting impurity lines.  The self-energy $\hat\Sigma(\varpi_l)$ due to impurities is
  \begin{align} \label{Sigma}
  \hat \Sigma(\varpi_l)&=n_i u_0^2\sum_p \hat G_{\bm p}(\varpi_l)
  =i\varpi_l [1-\eta_1(\varpi_l)] \hat\tau_0 - \eta_2(\varpi_l) \hat\tau_3.
  \end{align}
One can notice that the terms with $\hat\tau_1$ and $\hat\tau_2$ vanish after angular integration of $\Delta(\bm p)$ over $\bm p$ in Eq.~\eqref{Sigma}.  Here we introduced the functions $\eta_1(\varpi_l)$ and $\eta_2(\varpi_l)$ defined as
  \begin{align}
  \eta_1(\varpi_l) & = 1 + \frac{1}{\tau \sqrt{\varpi_l^2+\Delta_0^2}},
  \label{eta_1} \\
  \eta_2(\varpi_l) & = \frac {1}{\tau\pi} \int_{-\omega_D}^{\omega_D}
  \frac{N(\xi)  \, d\xi}{N(0)} \frac{\xi}{\varpi_l^2+\Delta_0^2+\xi^2},
  \label{eta_2} \\
  \frac{1}{\tau} & = n_i u_0^2 \pi N(0),
   \label{tau}
  \end{align}
where $\tau$ is the quasiparticle lifetime due to scattering on impurities.  One can notice that, within the Gaussian model of disorder, the concentration of impurities $n_i$ and the strength of the impurity potential $u_0$ can be completely absorbed into the definition of the quasiparticle lifetime $\tau$ in Eq.~\eqref{tau}.  This is not the case for the non-Gaussian model of disorder discussed in Sec.~\ref{sec:skew}.

The function $\eta_1(\varpi_l)$ in Eq.~\eqref{eta_1} is well known in the theory of impurity scattering.  Moreover, the first term with $\hat\tau_0$ in Eq.~\eqref{Sigma} is, essentially, the same as in Eq.~\eqref{loop}.  In contrast, the function $\eta_2(\varpi_l)$ in Eq.~\eqref{eta_2} is less common and deserves more discussion, because it will play crucial role in this Section.  The function $N(\xi)$ in Eq.~\eqref{eta_2} is the energy density of states as a function of the electron energy $\xi$.  If we make the approximation $N(\xi)=N(0)$, i.e.,\ take $N(\xi)$ to be a constant, then the integral in Eq.~\eqref{eta_2} vanishes, because the integrand is odd in $\xi$.  However, if we take into account that $N(\xi)$ is not a constant, i.e.,\ consider the particle-hole asymmetry, then $\eta_2(\varpi_l)$ is non-zero.  Let us expand $N(\xi)$ near the Fermi energy
  \begin{align} \label{N'}
  N(\xi) \approx N(0) + N'(0)\,\xi,
  \end{align}
where $N'(0)=dN/d\xi$ is taken at $\xi=0$.  When Eq.~\eqref{N'} is substituted into Eq.~\eqref{eta_2}, the integral diverges.  However, as we will see later, this divergence cancels out in the final results, and the outcome can be expressed in terms of $N'(0)$, which is a measure of the particle-hole asymmetry.

Using $\hat \Sigma(\varpi_l)$ from Eq.~\eqref{Sigma} to solve Dyson's equation
  \begin{align}
  \hat{\cal G}_{\bm p}^{-1}(\varpi_l) = \hat G^{-1}_{\bm p}(\varpi_l)
  - \hat\Sigma(\varpi_l),
  \end{align}
we obtain the dressed Green's function
  \begin{align} \label{eq:dressed_Green's}
  {\cal G}_{\bm p}(\varpi_l) \! = \!
  -\frac{i\varpi_l\eta_1\hat\tau_0
  \!+\! (\xi_p\!-\!\eta_2)\hat\tau_3
  \!+\!p_x\Delta_x \hat\tau_1\!-\!p_y\Delta_y \hat\tau_2}
  {\varpi_l^2\,\eta_1^2 + (\xi_p-\eta_2)^2 + \Delta_0^2 } \! .
  \end{align}

In order to calculate the two-particle response function self-consistently, it is necessary to include vertex corrections due to impurity lines.  They transform the
bare vertex $\hat\gamma_j(\bm p)=-ev_j(\bm p)\hat\tau_0$ of interaction with the electromagnetic field in Eq.~\eqref{S_em} into the dressed vertex $\hat\Gamma_j(\bm p)$.  The vertex $\hat\Gamma_j(\bm p)$ is obtained by solving the Bethe-Salpeter equation illustrated in Fig.~\ref{fig:self_vertex}b
  \begin{align} \label{Bethe-Salpeter}
  \hat\Gamma_j(\bm p) = \hat\gamma_j(\bm p) + n_iu_0^2 \sum_{\bm p'}
  \hat\tau_3 \hat{\cal G}_{\bm p'}(\varpi_l) \hat\Gamma_j(\bm p')
  \hat{\cal G}_{p'}(\varpi_l \!+\! \omega_n) \hat\tau_3.
  \end{align}
For elastic impurity scattering, the integral equation \eqref{Bethe-Salpeter} can be solved analytically by summing the geometric series as shown in Appendix~\ref{vertex_corr}.  The solution for $\hat\Gamma_y$, which contributes to the anomalous Hall response [see Eq.~\eqref{eq:dressed_bubble}], is given by Eq.~\eqref{Gamma}
  \begin{align}
  & \hat\Gamma_y(\bm p) = \gamma_y(\bm p) + \frac{\Delta_y p_F v_F}{2\pi\tau} \,
  (-i a_1 \hat\tau_1+ a_2 \hat\tau_2) \, L(\varpi_l,\omega_n)
  \nonumber \\
  & \times \left( \frac{1-b_1}{(1-b_1)^2+b^2_3} \hat\tau_0
  + \frac{i b_2}{(1-b_1)^2+b^2_3} \hat\tau_3 \right)
  \label{eq:fullvertex}.
  \end{align}
Here the functions $a_1$ and $a_2$ are defined as
  \begin{align}
  a_1(\varpi_l,\omega_n) &= \eta_2(\varpi_l)-\eta_2(\varpi_l + \omega_n),
  \label{eq:a1} \\
  a_2(\varpi_l,\omega_n) &= i\varpi_l \eta_1(\varpi_l)
  + i(\varpi_l + \omega_n) \, \eta_1(\varpi_l + \omega_n),
  \label{eq:a2}
\end{align}
and the function $L(\varpi_l,\omega_n)$ is
\begin{align}
  & L(\varpi_l,\omega_n)= \int_{-\omega_D}^{\omega_D} \frac{N(\xi)\, d\xi}{N(0)}
  \frac{1}{D(\varpi_l,\xi) \, D(\varpi_l\!+\!\omega_n,\xi)}
  \nonumber \\
  & \approx \!\frac{1}{(2\varpi_l\!+\!\omega_n)\omega_l}
  \!\left(\! \frac{1}{\sqrt{\Delta_0^2\!+\!\varpi_l^2}}
  \!-\!\frac{1}{\sqrt{\Delta_0^2+(\varpi_l\!+\!\omega_n)^2}} \!\right),
  \label{eq:L} \\
  & D(\varpi_l,\xi)=\varpi_l^2 \eta_1^2(\varpi_l) + [\xi-\eta_2(\varpi_l)]^2
  + \Delta_0^2.
  \nonumber
  \end{align}
The functions $b_1$ and $b_2$ are given by the integrals
  \begin{align}
  & b_1(\varpi_l,\omega_n) =  \frac{1}{\pi\tau} \int_{-\omega_D}^{\omega_D} \frac{N(\xi)\,d\xi}{N(0)}
  \label{eq:b1} \\
  & \times \frac{ [\xi\!-\!\eta_2(\varpi_l)][\xi\!-\!\eta_2(\varpi_l^+)]
  \!+\!\varpi_l\varpi_l^+\eta_1(\varpi_l)\eta_1(\varpi_l^+)}
  {D(\varpi_l,\xi) \, D(\varpi_l^+,\xi)},
  \nonumber \\
  \nonumber \\
  & b_2(\varpi_l,\omega_n) =  \frac{1}{\pi\tau} \int_{-\omega_D}^{\omega_D} \frac{N(\xi)\, d\xi}{N(0)}
  \label{eq:b2} \\
  & \times \frac{ \varpi_l^+\eta_1(\varpi_l^+) [\xi\!-\!\eta_2(\varpi_l)]
  \!-\!\varpi_l\eta_1(\varpi_l) [\xi\!-\!\eta_2(\varpi_l^+)]}
  {D(\varpi_l,\xi) \, D(\varpi_l^+,\xi)},
  \nonumber
  \end{align}
where $\varpi_l^+=\varpi_l+\omega_n$.

Using Eqs.~(\ref{eq:dressed_Green's}) and (\ref{eq:fullvertex}), one can now calculate the dressed response function ${\cal Q}^{(4)}_{xy}$ shown in Fig.~\ref{fig:self_vertex}c \cite{footnote3}
  \begin{align} \label{eq:dressed_bubble}
  {\cal Q}^{(4)}_{xy}(\omega_n) = {\rm Tr} \, [\hat\gamma_x(\bm p) \,
  \hat{\cal G}_{\bm p}(\varpi_l) \, \hat\Gamma_y(\bm p) \,
  \hat{\cal G}_{\bm p}(\varpi_l+\omega_n)].
  \end{align}
This expression can be simplified by integrating over the angle of $\bm p$ first. After substituting the dressed vertex $\hat\Gamma_y(\bm p)$, the contribution from the first term in Eq.~(\ref{eq:fullvertex}) drops out after taking the trace over the Nambu space, as discussed in Sec.~\ref{Sec:Clean}.  Thus, we are left only with the second, momentum-independent term in Eq.~(\ref{eq:fullvertex}), and the integral over $\bm p$ in Eq.~\eqref{eq:dressed_bubble} reduces to
  \begin{align}
  & \sum_{\bm p} \hat{\cal G}_{\bm p}(\varpi_l+\omega_n) \, \hat\gamma_x(\bm p)
  \, \hat{\cal G}_{\bm p}(\varpi_l)
  \nonumber \\
  & = \frac{\Delta_x p_F^2}{4\pi} \,
  (a_2 \hat\tau_1 - i a_1 \hat\tau_2) \, L(\varpi_l,\omega_n).
  \label{GgG}
  \end{align}
[One can notice that, in the zeroth order in disorder ($\eta_1=1$ and $\eta_2=0$), Eq.~\eqref{GgG} reduces to Eq.~\eqref{Lambda_x}.]  Substituting Eq.~\eqref{GgG} into Eq.~(\ref{eq:dressed_bubble}), i.e.,\ multiplying Eq.~\eqref{GgG} by the second term in Eq.~(\ref{eq:fullvertex}), and taking the trace, we obtain the response function
  \begin{align}
  {\cal Q}^{(4)}_{xy}(\omega_n) & =
  -\frac{e^2 \Delta_x \Delta_y p_F^3 v_F}{4\pi^2 \tau}
  T \sum_{\varpi_l} L^2(\varpi_l,\omega_n)
  \label{Q_xy} \\
  & \times \left( \frac{-4i a_1 a_2 (1-b_1)-2(a_1^2+a_2^2) b_2}
  {(1-b_1)^2+b_2^2} \right).
  \nonumber
  \end{align}
In contrast to the clean case discussed in Sec.~\ref{Sec:Clean}, the response function \eqref{eq:dressed_bubble} does not vanish identically, because the second term in Eq.~(\ref{eq:fullvertex}) is momentum-independent as a consequence of randomization of electron momentum due to scattering on impurities.  However, if the model has the particle-hole symmetry, then the functions $a_1$~\eqref{eq:a1} and $b_2$~\eqref{eq:b2} vanish, and, thus, the anomalous Hall response \eqref{Q_xy} is zero to any order in impurity scattering.  To obtain a non-zero result, we need to take into account the particle-hole asymmetry explicitly, as discussed in the next Subsection.

\subsection{Perturbative expansion in the strength of disorder}

As mentioned in Sec.~\ref{Sec:disorder}, $\rm Sr_2RuO_4$ is a stoichiometric  crystalline material with a very low concentration of impurities, and the existence of $p$-wave pairing requires that $\Delta_0\gg1/\tau$.  Thus, we can use the strength of disorder as a small parameter and expand the anomalous Hall response function \eqref{Q_xy} to the lowest non-vanishing order in $1/\tau$.  The non-vanishing term appears in the order $(1/\tau)^2$, i.e.,\ in the fourth order in $V_{\rm imp}$:
  \begin{align}
  Q^{(4)}_{xy}(\omega_n) & = -s_{xy}3 \frac{e^2 \Delta_0^2 p_F v_F}{2\pi^2 \tau}
  T \sum_{\varpi_l} L^2(\varpi_l,\omega_n)
  \label{Q^4} \\
  & \times (2\varpi_l+\omega_n) \, [\eta_2(\varpi_l)-\eta_2(\varpi_l+\omega_n)].
  \nonumber
  \end{align}
Notice that one power of $1/\tau$ appears in the prefactor in Eq.~\eqref{Q^4}, and another power comes from the function $\eta_2$ defined in Eq.~\eqref{eta_2}.  Eq.~\eqref{Q^4} precisely corresponds to the diagrams shown in Fig.~\ref{fig:2nd_order_line}.

Using Eqs.~\eqref{eta_2} and \eqref{N'}, the difference $\eta_2(\varpi_l)-\eta_2(\varpi_l+\omega_n)$ in Eq.~\eqref{Q^4} can be written as
  \begin{align}
  \eta_2(\varpi_l) & -\eta_2(\varpi_l+\omega_n) =
  \frac{1}{\tau \pi }\int_{-\omega_D}^{\omega_D} \frac{N(\xi) d\xi}{N(0)}
  \label{eta-eta} \\
  & \times \xi \left( \frac{1}{\varpi_l^2+\xi^2+\Delta_0^2}
  -\frac{1}{(\varpi_l+\omega_n)^2+\xi^2+\Delta_0^2} \right)
  \nonumber \\
  & = \frac{1}{\tau} \frac{N'(0)}{N(0)} \left(
  \sqrt{(\omega_n+\varpi_l)^2+\Delta_0^2}-\sqrt{\varpi_l^2+\Delta_0^2} \right)
  \nonumber \\
  & -\frac{2}{\tau\pi} \frac{N'(0)}{N(0)}
  \frac{\omega_n(2\varpi_l+\omega_n)}{\omega_D}.
  \label{eq:ph_asymmetry}
  \end{align}
When Eq.~\eqref{N'} is substituted into Eq.~\eqref{eta-eta}, the constant term $N(0)$ in the density of states cancels out, and the term proportional to  $N^\prime(0)$ produces a convergent integral over $\xi$.  Thus, the anomalous Hall response \eqref{Q^4} becomes directly proportional to the particle-hole asymmetry parameter $N^\prime(0)$ and would vanish if $N^\prime(0)=0$.

For generality, we consider the finite limits of integration $\pm\omega_D$ in Eq.~\eqref{eta-eta}.  The first term in Eq.~\eqref{eq:ph_asymmetry} represents the value of the integral when the limits are taken to $\pm\infty$, whereas the second term is  a correction due to the finite limits of integration, assuming that $\omega_D\gg\Delta_0,\omega_n$.  As we will see below, one of these two terms gives the dominant contribution to the imaginary, and the other one to the real part of the ac Hall conductivity in the high-frequency limit.  We first evaluate the cutoff-independent contribution to $\sigma_{xy}(\omega)$ given by the first term in Eq.~\eqref{eq:ph_asymmetry} and then the cutoff-dependent contribution from the second term.

\subsubsection{Cutoff-independent contribution to the anomalous Hall conductivity}
\label{Sec:Cutoff-independent}

Substituting the first term in Eq.~\eqref{eq:ph_asymmetry} into Eq.~\eqref{Q^4} (or taking the limit $\omega_D\to\infty$), we obtain the cutoff-independent contribution  $Q^{(4a)}_{xy}$ to the anomalous Hall response function
  \begin{align} \label{Q^4a}
  Q^{(4a)}_{xy}(\omega_n) = -s_{xy} \frac{3e^2\nu\Delta_0}{\pi^2\tau^2\omega_n^2}
  K(\omega_n),
  \end{align}
where the dimensionless parameter $\nu$ is a measure of the particle-hole asymmetry
  \begin{align} \label{nu}
  \nu = \frac{p_F v_F N'(0)}{2 N(0)},
  \end{align}
and the dimensionless function $K(\omega_n)$ is
  \begin{align} \label{eq:K0}
  K(\omega_n) \!& \!=\! \sum_{\varpi_l}\frac{T\Delta_0}{2\varpi_l+\omega_n}
  \!\!\left( \frac{1}{\sqrt{\varpi_l^2\!+\!\Delta_0^2}}
  \!-\!\frac{1}{\sqrt{(\varpi_l\!+\!\omega_n)^2\!+\!\Delta_0^2}}\right)^2
  \nonumber \\
  & \times \left( \sqrt{(\omega_n+\varpi_l)^2+\Delta_0^2}
  -\sqrt{\varpi_l^2+\Delta_0^2} \right).
  \end{align}
The Matsubara sum in Eq.~\eqref{eq:K0} can be evaluated using complex analysis as shown in Appendix~\ref{app:calc_K}.  The function $K(\omega_n)$ can be written as the sum of three terms
  \begin{align} \label{eq:K}
  K(\omega_n) = K_1(\omega_n) + K_{2a}(\omega_n) + K_{2b}(\omega_n),
  \end{align}
where the functions $K_1(\omega_n)$, $K_{2a}(\omega_n)$, and $K_{2b}(\omega_n)$ are given by Eqs.~\eqref{eq:K22}, \eqref{eq:K11}, and \eqref{eq:K33}.  After the analytical continuation \eqref{Omega+}, we obtain
  \begin{align}
  & K_{1}(\omega) = \int\limits_1^\infty \frac{dx}{2\pi}
  \frac{-i 12 \, \tilde\omega}{\sqrt{x^2-1} \, (\tilde\omega_+^2-4x^2)}
  \tanh\left(\frac{x\Delta_0}{2T}\right) \!,
  \label{K_1} \\
  & K_{2a}(\omega) \! = \! \frac{1}{2} \tanh\left(\frac{\Delta_0}{2T}\right) \!\!
  \left[ \sqrt{\frac{\tilde\omega^+}{\tilde\omega^++2}}
  + \sqrt{\frac{\tilde\omega^+}{\tilde\omega^+-2}} \right] \!,
  \label{K_2a} \\
  & K_{2b}(\omega) = 2i \int_1^\infty \frac{dx}{2\pi} \sqrt{x^2-1}
  \tanh\left(\frac{x\Delta_0}{2T}\right)
  \label{K_2b} \\
  \!&\! \times \!\! \left[\!
  \frac{1}{(2x\!+\!\tilde\omega^+)(1-[x\!+\!\tilde\omega^+]^2)}
  \!-\!\frac{1}{(2x\!-\!\tilde\omega^+)(1-[x\!-\!\tilde\omega^+]^2)} \!\right] \!,
  \nonumber
  \end{align}
where $\tilde\omega^+=(\omega+i\delta)/\Delta_0$.

Using Eqs.~\eqref{Q^4a} and \eqref{eq:sigma-q}, we find the cutoff-independent contribution $\sigma^{(4a)}_{xy}$ to the anomalous Hall conductivity
  \begin{align} \label{sigma^4a}
  \sigma^{(4a)}_{xy}(\omega) = -s_{xy} \frac{e^2}{\hbar}
  \frac{3\nu\Delta_0}{\pi^2\tau^2\omega^3}
  \, i K\left(\frac{\omega^+}{\Delta_0}\right),
  \end{align}
where the dimensionless function $K$ \eqref{eq:K} is given by the sum of Eqs.~\eqref{K_1}, \eqref{K_2a}, and \eqref{K_2b}.

\begin{figure} \centering
\includegraphics[width=0.99\linewidth]{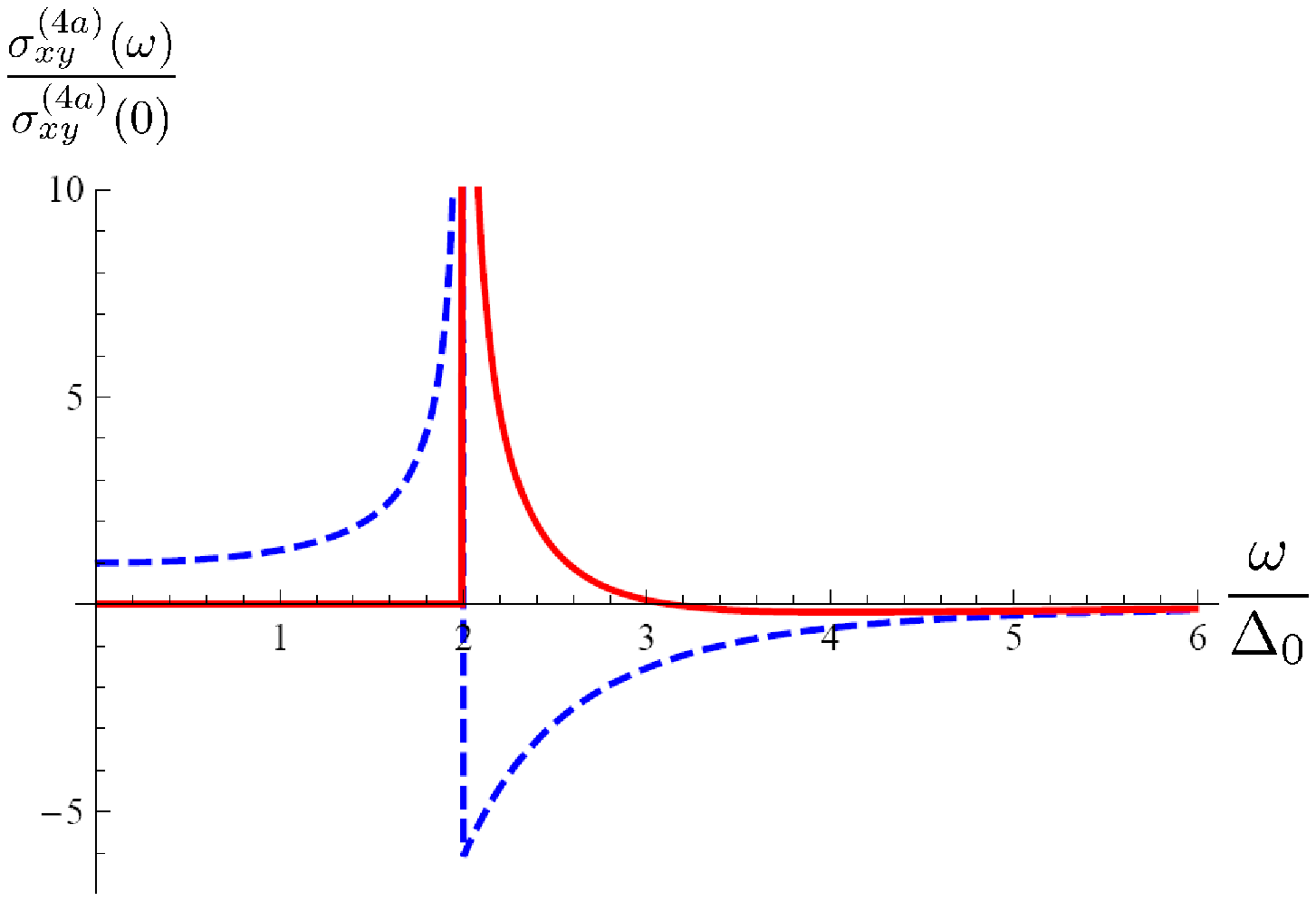}
  \caption{(Color online) Real (dashed line) and imaginary (solid line) parts of the cutoff-independent contribution $\sigma^{(4a)}_{xy}(\omega)$ to the anomalous Hall conductivity at $T=0$ given by Eqs.~\eqref{sigma^4a} and \eqref{K(T=0)}. } \label{fig:S}
\end{figure}

At $T=0$, the function $K(y)$ becomes
  \begin{align} \label{K(T=0)}
  & K(y) = \frac12 \left(\sqrt{\frac{y}{y+2}}+\sqrt{\frac{y}{y-2}} \right)
  + \frac{6i}{\pi} \frac{\arcsin(y/2)}{\sqrt{4-y^2}}
  \\
  & +i \int\limits_1^{\infty} \frac{dx}{\pi} \!\!
  \left[\!\frac{\sqrt{x^2-1}}{(2x\!+\!y)(1-[x\!+\!y]^2)}
  \!-\!\frac{\sqrt{x^2-1}}{(2x\!-\!y)(1-[x\!-\!y]^2)} \! \right] \! .
  \nonumber
  \end{align}
The plots of the real and imaginary parts of $\sigma^{(4a)}_{xy}$ vs.\ $\omega$ at $T=0$, obtained from Eqs.~\eqref{sigma^4a} and \eqref{K(T=0)}, are shown in Fig.~\ref{fig:S}.  At $y=2$, the function $K(y)$ exhibits a square-root singularity.  The asymptotes of $K(y)$ are
  \begin{align}
  - iK(y)= \left\{ \begin{array}{ll}
  \displaystyle  \frac{1}{15\pi} \, y^3, & \quad y \rightarrow 0
  \\ \\
  \displaystyle - i - \frac{6}{\pi y}\ln y, & \quad y \rightarrow \infty.
  \end{array}  \right.
  \end{align}
So, at high frequencies $\omega\gg\Delta_0$, $\sigma^{(4a)}_{xy}$ becomes
  \begin{align}
  \sigma^{(4a)}_{xy}(\omega) = - s_{xy} \frac{e^2}{\hbar}
  \frac{3\nu\Delta_0}{\pi^2\tau^2\omega^3}
  \left[i + \frac{6}{\pi} \frac{\Delta_0}{\omega}
  \ln\left(\frac{\omega}{\Delta_0}\right) \right],
  \end{align}
and the dc limit is
  \begin{align} \label{sigma^4a_dc}
  \sigma^{(4a)}_{xy}(\omega=0) = s_{xy} \frac{e^2}{\hbar} \frac{\nu}{5\pi^3}
  \left( \frac{1}{\tau\Delta_0} \right)^2 .
  \end{align}
The electron-hole asymmetry factor $\nu$ given by Eq.~\eqref{nu} is, generally, of the order of unity, whereas the factor $\tau\Delta_0\gg1$  in Eq.~\eqref{sigma^4a_dc} is large, as required for the existence of $p$-wave superconductivity.  Thus, the anomalous dc Hall conductivity \eqref{sigma^4a_dc} per layer is always much smaller than the quantum of conductance $e^2/h$.

\subsubsection{Cutoff-dependent contribution to the anomalous Hall conductivity}
\label{Sec:Cutoff-dependent}

Now we consider the contribution $Q^{(4b)}_{xy}$ to the anomalous Hall response function \eqref{Q^4} from the second term in Eq.~\eqref{eq:ph_asymmetry}, which depends on the energy cutoff $\omega_D$:
  \begin{align}  \label{eq:Q4b}
  Q^{(4b)}_{xy}(\omega_n) = -s_{xy}
  \frac{6 e^2 \nu \Delta_0}{\pi^3 \tau^2 \omega_D \omega_n} J(\omega_n),
  \end{align}
where the dimensionless function $J(\omega_n)$ is
  \begin{align}  \label{eq:J0}
  J(\omega_n) = T \Delta_0 \sum_l \left( \frac{1}{\sqrt{\varpi_l^2\!+\!\Delta^2}}
  \!-\!\frac{1}{\sqrt{(\varpi_l\!+\!\omega_n)^2\!+\!\Delta^2}} \right)^2.
  \end{align}
The Matsubara sum in Eq.~(\ref{eq:J0}) is calculated in Appendix~\ref{app:calc_J}. The result for $J(\omega_n)$ is
  \begin{align}  \label{eq:J}
  & J(\omega_n) = \tanh\left(\frac{\Delta_0}{2T}\right)
  - \frac{2}{\pi} \int\limits^\infty_{\Delta_0}
  \frac{dx}{\sqrt{x^2-\Delta_0^2}} \tanh\left(\frac{x}{2T} \right)
  \nonumber \\
  & \times \left( \frac{\Delta_0}{\sqrt{(ix+\omega_n)^2+\Delta_0^2}}
  + \frac{\Delta_0}{\sqrt{(ix-\omega_n)^2+\Delta_0^2}} \right).
  \end{align}
Performing the analytical continuation \eqref{Omega+} in Eqs.~\eqref{eq:J}, \eqref{eq:Q4b}, and \eqref{eq:sigma-q}, we obtain the cutoff-dependent term $\sigma^{(4b)}_{xy}$ in the anomalous Hall conductivity:
  \begin{align}  \label{sigma^4b}
  \sigma^{(4b)}_{xy}(\omega) = s_{xy} \frac{e^2}{\hbar}
  \frac{6 \nu \Delta_0}{\pi^3 \tau^2 \omega_D \omega^2}
  J\left(\frac{\omega^+}{\Delta_0}\right).
  \end{align}
Here the dimensionless function $J(y)$ is
  \begin{align} \label{J}
  J(y) & = \tanh\left(\frac{\Delta_0}{2T}\right) - \frac{2}{\pi}
  \int^\infty_1 \frac{dx}{\sqrt{x^2 - 1}} \tanh\left(\frac{x\Delta_0}{2T}\right)
  \nonumber \\
  & \times \left( \frac{1}{\sqrt{1-(x-y)^2}}
  + \frac{1}{\sqrt{1-(x+y)^2}} \right) .
  \end{align}

\begin{figure} \centering
\includegraphics[width=0.9\linewidth]{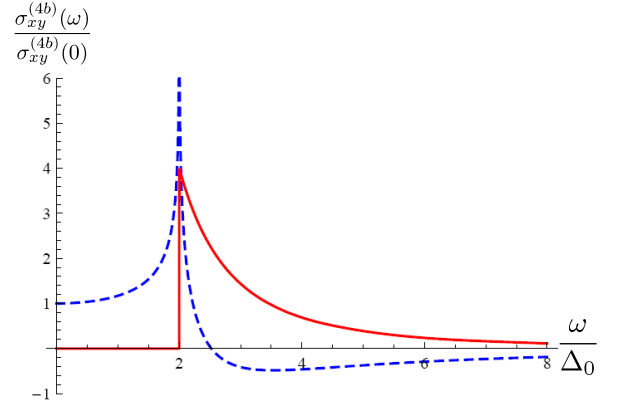}
  \caption{(Color online) Real (dashed line) and imaginary (solid line) parts of the cutoff-dependent contribution $\sigma^{(4b)}_{xy}(\omega)$ to the anomalous Hall conductivity at $T=0$ given by Eqs.~\eqref{sigma^4b} and \eqref{J(T=0)}. } \label{fig:G}
\end{figure}

At $T=0$, the explicit analytical expression for the function $J(y)$ can be obtained from Eq.~\eqref{J}:
  \begin{align} \label{J(T=0)}
  & J(y) = 1 - \frac{2}{\pi} \int\limits^\infty_1 \frac{dx}{\sqrt{x^2\!-\!1}}
  \!\left(\! \frac{1}{{\sqrt{1\!-\!(x\!-\!y)^2}}}\!
  + \!\frac{1}{{\sqrt{1\!-\!(x\!+\!y)^2}}} \!\right)
  \nonumber\\
  & = - 1 - \frac{4}{\pi\sqrt{4-y^2}} \left[ F\!
  \left(i{\rm Arcsinh}\!\left\{\sqrt{\frac{2-y}{y}}\right\};
  \frac{y^2}{y^2\!-\!4}\right) \right.
  \nonumber \\
  & \left. -F\! \left(i{\rm Arcsinh}\!\left\{\sqrt{-\frac{2+y}{y}}\right\};
  \frac{y^2}{y^2\!-\!4}\right) \right].
  \end{align}
The function $F(x;k)$ is the incomplete elliptic integral of the first kind.  The plots of the real and imaginary parts of $\sigma^{(4b)}_{xy}$ vs.\ $\omega$ at $T=0$, obtained from Eqs.~\eqref{sigma^4b} and \eqref{J(T=0)}, are shown in
Fig.~\ref{fig:G}.  At $y=2$, the imaginary part of $J(y)$ has a discontinuity (jump), whereas the real part of $J(y)$ has a logarithmic divergence.  The function $J(y)$ has the following asymptotes:
  \begin{align}
  J(y)= \left\{
  \begin{array}{ll}
  \displaystyle - \frac{1}{16} \, y^2, & \quad y\to0
  \\ \\
  \displaystyle 1 - \frac{4i}{\pi y}\ln y, & \quad y\to\infty
  \end{array}.  \right.
  \end{align}
So, at high frequencies $\omega\gg\Delta_0$, $\sigma^{(4b)}_{xy}$ becomes
  \begin{align}  \label{Hall_phasym_high}
  \sigma^{(4b)}_{xy}(\omega) = s_{xy} \frac{e^2}{\hbar}
  \frac{6 \nu \Delta_0}{\pi^3 \tau^2 \omega_D \omega^2}
  \left[ 1 - \frac{4i}{\pi} \frac{\Delta_0}{\omega}
  \ln \left(\frac{\omega}{\Delta_0}\right) \right].
  \end{align}
and the dc limit is
  \begin{align}  \label{sigma^4b_dc}
  \sigma^{(4b)}_{xy}(\omega=0) = - s_{xy} \frac{e^2}{\hbar}
  \frac{6 \nu}{16 \pi^3 \tau^2 \omega_D \Delta_0}.
  \end{align}

\subsubsection{Comparison of the cutoff-dependent and -independent contributions to the anomalous Hall conductivity}

Combining Eqs.~\eqref{sigma^4a} and \eqref{sigma^4b}, we obtain the total anomalous Hall conductivity $\sigma_{xy}^{(4)}$ within the Gaussian disorder model with particle-hole asymmetry
  \begin{align}  \label{eq:sigma4}
  \sigma_{xy}^{(4)} = \sigma^{(4a)}_{xy} + \sigma^{(4b)}_{xy}.
  \end{align}

At high frequencies $\omega\gg\Delta_0$, the asymptotic expressions for the real and imaginary parts of the anomalous Hall conductivity \eqref{eq:sigma4} are
  \begin{align} \label{eq:sigma4_real}
  \sigma'^{(4)}_{xy}(\omega) & = s_{xy} \frac{e^2}{\hbar}
  \frac{3\nu\Delta_0}{\pi^3\tau^2} \!
  \left[\frac{2}{\omega_D \omega^2} - \frac{6\Delta_0}{\omega^4}
  \ln\left(\frac{\omega}{\Delta_0}\right) \right] \! ,
  \\
  \label{eq:sigma4_imag}
  \sigma''^{(4)}_{xy}(\omega) &  =  s_{xy} \frac{e^2}{\hbar}
  \frac{3\nu\Delta_0}{\pi^2\tau^2} \!
  \left[-\frac{1}{\omega^3} - \frac{8\Delta_0}{\pi^2\omega_D \omega^3}
  \ln\left(\frac{\omega}{\Delta_0}\right) \right] \! .
  \end{align}
Eq.~\eqref{eq:sigma4_imag} shows that the dominant contribution to the imaginary part $\sigma''^{(4)}_{xy}$ is given by the cutoff-independent term.  In contrast, Eq.~\eqref{eq:sigma4_real} shows that, in the case of $\Delta_0 \ll \omega \ll \omega_D \ll \omega^2/[\Delta_0\ln(\omega/\Delta_0)]$, the dominant contribution to the real part $\sigma'^{(4)}_{xy}$ is given by the cutoff-dependent term.  This is why it is important to perform calculations keeping the cutoff $\omega_D$ large, but finite.

On the other hand,  the low-frequency anomalous Hall conductivity is dominated by the cutoff-independent term: $\sigma^{(4a)}_{xy}\gg\sigma^{(4b)}_{xy}$ from Eqs.~\eqref{sigma^4a_dc} and \eqref{sigma^4b_dc} at $\omega=0$.

\section{Experimental implications} \label{sec:PKE}

\subsection{Estimates of the anomalous Hall conductivity}

In this Subsection, we discuss experimental implications of our results
(\ref{eq:sigma3}) and~(\ref{eq:sigma4}) for the anomalous Hall conductivity.  First, we estimate the magnitude of the anomalous dc Hall conductivity $\sigma_{xy}(\omega=0)$.

Within the Gaussian disorder model, the dc Hall conductivity is given by Eq.~\eqref{sigma^4a_dc}.  [As discussed in the preceding Section, Eq.~\eqref{sigma^4b_dc} gives a negligible contribution compared with Eq.~\eqref{sigma^4a_dc}.]  For a crude estimate, we use the following values of the parameters in Eq.~\eqref{sigma^4a_dc}: $\Delta_0=0.23$~meV, as estimated in Sec.~\ref{Sec:Impurities} from the value of $T_c$, $1/\tau\approx 7\times10^{-5}$ eV taken from Ref.~\cite{xia_prl'06}, and $\nu\sim10$.  The particle-hole asymmetry parameter $\nu$ \eqref{nu}, in principle, can be deduced more accurately from the band-structure calculations \cite{Mazin_PhysRevB'00} for $\rm Sr_2RuO_4$.  Thus we obtain an estimate for $\sigma^{(4)}_{xy}$ from Eq.~\eqref{sigma^4a_dc}
  \begin{align}
  \sigma^{(4)}_{xy}(\omega=0) & \sim  10^{-2}\, \frac{e^2}{\hbar}.
  \label{dc_estimate_4b}
  \end{align}
Because $\tau\Delta_0\gg1$, the dc Hall conductivity \eqref{sigma^4a_dc} within the Gaussian disorder model is much smaller than the conductance quantum $e^2/h$.

Now we consider the anomalous dc Hall conductivity $\sigma^{(3)}_{xy}(\omega=0)$ originating from the non-Gaussian disorder model.  Eq.~\eqref{Hall_skew_dc} for $\sigma^{(3)}_{xy}(\omega=0)$ contains the parameter $W$~\eqref{W} proportional to the combination $\kappa_3 n_i u_0^3$, which is different from the combination $n_i u_0^2$ appearing in Eq.~\eqref{tau} for $\tau$.  Thus, the knowledge of the impurity scattering time $\tau$ alone is not sufficient to calculate $W$.  In addition to $\tau$, one needs to know the concentration of impurities $n_i$ and the skewness parameter $\kappa_3$.  Goryo~\cite{goryo_prb'08} estimated the typical distance between impurities as $l_i=1000-5000$~\AA.  For the estimates in our paper, we use the value $l_i=1000$~\AA, which corresponds to $n_i\sim10^{14}$ m$^{-2}$ \cite{footnote_Goryo}.  Using Eq.~\eqref{N(0)} for $N(0)$ and Eq.~\eqref{tau} for $1/\tau$, we rewrite the expression for $W$ in the following form
\begin{align} \label{W-estimate}
  |W| = \kappa_3\,\sqrt{\frac{p_F^3 \hbar v_F}{32 n_i \tau^3}}
  \approx (6 \; \rm meV)^2.
\end{align}
For the numerical estimate given in Eq.~\eqref{W-estimate}, we used the values $p_F=0.75$~\AA$^{-1}$ and $v_F=5.5\times10^4$~m/s from Ref.~\cite{Mackenzie_RevModPhys'03} (giving $E_F=p_Fv_F/2=0.14$~eV) and the values of $1/\tau$ and $n_i$ quoted above.  For an estimate of the skewness parameter, we took the upper limit $\kappa_3=1$, not having information about the actual nature of impurities.  This value corresponds, for example, to randomly distributed impurities generating short-range, delta-function potentials of the equal strength $u_0$, as discussed after Eq.~\eqref{3rd-moment}.  The same assumption was utilized in Ref.~\cite{goryo_prb'08}.

Substituting Eq.~\eqref{W-estimate} into Eq.~\eqref{Hall_skew_dc}, we obtain an estimate for the magnitude of the dc Hall conductivity within the non-Gaussian model
  \begin{align}
  \sigma^{(3)}_{xy}(\omega=0) \approx 18 \, \frac{e^2}{\hbar}.
  \label{sigma_3_dc}
  \end{align}
Eq.~\eqref{sigma_3_dc} gives a value much greater than the conductance quantum $e^2/h$, even though $\tau\Delta_0\gg 1$.  Thus, for the parameters given above, we find that the contribution from the skew-scattering diagrams to $\sigma_{xy}(\omega=0)$ is dominant.  This is because the skew-scattering processes correspond to the lower-order diagrams in impurity concentration.  Finally, we emphasize that our calculation is done for a single domain of the $p_x\pm ip_y$ superconductor, while a realistic macroscopic sample of $\rm Sr_2RuO_4$ should consists of multiple domains with opposite chiralities.  The dc Hall effect contributions of the opposite signs from different domains would cancel out and make an experimental verification of Eq.~\eqref{sigma_3_dc} difficult.

We now calculate the anomalous ac Hall conductivity, given by Eqs.~\eqref{eq:sigma3}, \eqref{sigma^4a}, \eqref{sigma^4b} and \eqref{eq:sigma4}, at the optical frequency $\omega=0.8$~eV utilized in the experiment~\cite{xia_prl'06}.  Some of these equations contain the cutoff frequency $\omega_D$, which depends on the microscopic nature of the pairing interaction and is not known very well.  In our calculations, we assumed that $\omega_D>\omega$.  Using the optical frequency $\omega$ as the lower bound for the cutoff frequency $\omega_D$, we find
  \begin{align} \label{eq:estimates_high-3}
  \sigma^{(3)}_{xy}(\omega=0.8\:{\rm eV}) &
  \sim \left(10^{-10} + 10^{-8} i\right) \, \frac{e^2}{\hbar},
  \\
  \sigma^{(4)}_{xy}(\omega=0.8\:{\rm eV}) &
  \sim  \left(10^{-12} + 10^{-12} i\right) \, \frac{e^2}{\hbar}.
  \label{eq:estimates_high-4}
  \end{align}
Comparing Eqs.~\eqref{eq:estimates_high-3} and \eqref{eq:estimates_high-4}, we observe that the non-Gaussian model gives a much greater contribution than the Gaussian model to both real and imaginary parts of $\sigma_{xy}(\omega=0.8\:{\rm eV})$.  Thus, in the rest of this Section, we will use Eq.~\eqref{eq:estimates_high-3} for the non-Gaussian model to estimate the Kerr angle.  However, one should keep in mind that the dominance of $\sigma^{(3)}_{xy}$ is the consequence of the high estimate for the parameter $W$ in Eq.~\eqref{W-estimate}.  This estimate depends on the parameters $\kappa_3$, $n_i$, and $u_0$, for which there are no direct measurements.  Similarly, $\sigma^{(4)}_{xy}$ for the Gaussian model depends on the parameters $\nu$ and $\omega_D$, for which there are no direct measurements either.  The parameter $\tau$, utilized for both models, is only indirectly inferred from the dc conductivity in Ref.~\cite{xia_prl'06}.  Given these great uncertainties, our numerical estimates should be considered as only tentative.

\subsection{Estimates of the polar Kerr angle}
\label{Sec:Kerr-estimate}

The anomalous Hall conductivity $\sigma_{xy}$ contributes to the
dielectric permeability tensor $\tensor\varepsilon$, which characterizes propagation of electromagnetic waves in the medium.  The dielectric permeability tensor $\tensor\varepsilon$ is related to the conductivity tensor $\tensor\sigma$ as follows
  \begin{align} \label{epsilon}
  \tensor\varepsilon =\tensor\varepsilon_{\infty}
  +\frac{4\pi i}{\omega} \, \tensor{\sigma}.
  \end{align}
Here $\tensor\varepsilon_{\infty}$ is the background dielectric tensor, which originates from polarizability of the other, non-conduction bands in the material.  The diagonal components of the conductivity tensor are assumed to have the Drude-like form:
  \begin{align}
  \sigma_{jj}=- \frac{\omega_j^2}{4\pi i (\omega+i\gamma_j)},
  \end{align}
where $\omega_j$ and $\gamma_j$ are the plasma frequency and the quasiparticle scattering rate along $j$-th axes, respectively.  Due to the square symmetry of $\rm Sr_2RuO_4$ in the $ab$-plane, we have $\omega_{x}=\omega_y=\omega_{ab}$, $\gamma_x=\gamma_y=\gamma_{ab}$, and $\varepsilon_{xx}=\varepsilon_{yy}=\varepsilon_{ab}$, so
  \begin{align} \label{epsilon_ab}
  \varepsilon_{ab}(\omega)=\varepsilon_{\infty}-\frac{\omega_{ab}^2}
  {\omega(\omega+i\gamma_{ab})}.
  \end{align}
It should be emphasized that we distinguish between $1/\tau$, the scattering rate on impurities given by Eq.~\eqref{tau}, and $\gamma$, the quasiparticle scattering rate at optical frequencies, which may be dominated by other scattering processes.

Now let us consider a polarized electromagnetic plane wave incident in the $z$ direction normal to the $(x,y)$ surface of a Q2D chiral superconductor, as relevant for the experiment~\cite{xia_prl'06}.  Propagation of the electromagnetic wave in the medium is described by Maxwell's equation:
  \begin{align}
  (c^2\partial_z^2 + \omega^2\tensor\varepsilon) \, \bm E=0,
  \end{align}
where $\bm E$ is polarized in the $(x,y)$ plane.  The anomalous Hall conductivity $\sigma_{xy}$ produces antisymmetric off-diagonal matrix elements $\varepsilon_{xy}=-\varepsilon_{yx}$ in the tensor $\tensor\varepsilon$ via Eq.~\eqref{epsilon}.  Thus, propagation of the wave inside of the superconductor is described by two circularly polarized eigenmodes with different refraction indices $n_{+}$ and $n_{-}$.  Given that the off-diagonal elements are typically very small $|\varepsilon_{xy}| \ll |\varepsilon_{xx}|$, one can expand $n_{+}$ and  $n_{-}$ to the first order in $\varepsilon_{xy}\propto\sigma_{xy}$.  Then, using the boundary conditions for the electric field at the interface between  vacuum and the material, one finds the reflection coefficient $|r|$ and the polar Kerr angle $\theta_K$ \cite{White-Geballe,mineev_prb'07}
  \begin{align}
  |r|&=\frac{|n-1|}{|n+1|},
  \label{r} \\
  \theta_K&=\frac{4\pi}{\omega d}\,{\rm Im}\left[\sigma_{xy}(\omega)\,
  \alpha(\omega)\right],
  \label{theta_K} \\
  \alpha(\omega)=\frac{1}{n(n^2-1)}&=\frac{1}{\sqrt{\varepsilon_{ab}
  (\omega)}\,[\varepsilon_{ab}(\omega)-1]}.
  \label{alpha}
  \end{align}
Here $n(\omega)=\sqrt{\varepsilon_{ab}(\omega)}$ is the complex refraction coefficient.  In Eq.~\eqref{theta_K}, $\sigma_{xy}(\omega)$ is the 2D Hall conductivity per one layer calculated in the previous Sections, and the interlayer distance $d=6.8$~\AA~\cite{Mackenzie_RevModPhys'03} converts it into the three-dimensional, bulk conductivity implied in Eq.~\eqref{epsilon}.  As Eq.~\eqref{theta_K} shows, the magnitude of the Kerr angle $\theta_K$ depends not only on the anomalous Hall conductivity $\sigma_{xy}(\omega)$, but also on the complex refraction coefficient $n(\omega)$.  The frequency dependence of the refraction coefficient is beyond the scope of the present paper, in which we focus on the TRSB effects. Therefore, we estimate the parameters determining $n(\omega)$ from the experimental data~\cite{Katsufuji_PRL'96}, where the out-of-plane optical conductivity of $\rm Sr_2RuO_4$ was studied. Because many parameters (e.g., the quasiparticle scattering rate $\gamma$) for the diagonal component $\sigma_{xx}(\omega)$ of the conductivity tensor are difficult to estimate, it is desirable to measure $n(\omega)$ experimentally for the same samples where the Kerr effect is measured.

According to Ref.~\cite{Katsufuji_PRL'96}, the dielectric constant and the plasma frequency in the $ab$-plane are $\varepsilon_{\infty}=10$ and $\omega_{ab}=2.9$~eV \cite{footnote4}.  Substituting these values into Eq.~\eqref{epsilon_ab}, we find the frequency of the plasma edge $\omega_p=\omega_{ab}/\sqrt{\varepsilon_{\infty}}=0.9$ eV.  So, the measurement frequency $\omega=0.8$~eV in the experiment \cite{xia_prl'06} appears to be below the plasma edge frequency $\omega_p$.  Another important parameter here is the quasiparticle scattering rate $\gamma_{ab}$.  There is experimental evidence that $\gamma_{ab}$ is quite large, of the order a fraction of eV~\cite{Katsufuji_PRL'96}.  To illustrate the importance of this parameter for the estimate of $\theta_K$, below we discuss two limits: $\omega\gg\gamma_{ab}$ and $\omega\sim\gamma_{ab}$.

The case of $\omega\gg\gamma_{ab}$ was previously considered by Goryo~\cite{goryo_prb'08}, who obtained the value for the Kerr angle to be about $30$~nrad.  However, as we showed in Sec~\ref{sec:skew}, the analytical structure of the Hall conductivity obtained by Goryo is incorrect, which leads to a much smaller estimate for the Kerr angle as we discuss below.  Indeed, consider the expression \eqref{theta_K} for the Kerr angle in the following form:
  \begin{align} \label{eq:thetaK1}
  \theta_K = \frac{4\pi}{\omega d} \left[ \alpha''(\omega)
  \sigma'_{xy}(\omega)+\alpha'(\omega) \sigma''_{xy}(\omega) \right],
  \end{align}
where $\alpha'(\omega)$ and $\alpha''(\omega)$ are the real and imaginary parts of $\alpha(\omega)$.  For the frequency $\omega<\omega_p$ below the plasma edge, Eq.~\eqref{epsilon_ab} gives $\varepsilon_{ab}(\omega)<0$, so the index of refraction $n(\omega)=\sqrt{\varepsilon_{ab}(\omega)}$ and $\alpha(\omega)$ in Eq.~\eqref{alpha} are imaginary.  Therefore, the main contribution to $\theta_K$ in Eq.~\eqref{eq:thetaK1} comes from real part of $\sigma_{xy}$.  Using the estimate $\sigma'_{xy}(\omega = 0.8\:{\rm eV})\sim 10^{-10} e^2/\hbar$ from Eq.~\eqref{eq:estimates_high-3}, we find $\theta_K \sim 0.1 $~nrad.  This estimate is much smaller than the one obtained by Goryo, because the real part of $\sigma_{xy}$ is three orders of magnitude smaller than imaginary part that was used in his estimate.

\begin{figure} \centering
\includegraphics[width=0.9\linewidth]{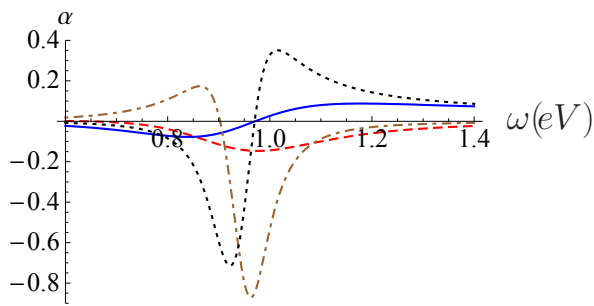}
  \caption{(Color online) Real and imaginary parts of the dimensionless function $\alpha(\omega)$ given by Eqs.~\eqref{alpha} and \eqref{epsilon_ab}.  The solid (blue) and dotted (black) lines represent $\alpha'(\omega)$ for $\gamma=0.4$ and $0.1$~eV, respectively.  The dashed (red) and dash-dot (brown) lines represent $\alpha''(\omega)$ for $\gamma=0.4$ and $0.1$~eV, respectively.  The parameters $\omega_{ab}=2.9$~eV and $\varepsilon_{\infty}=10$ are taken from Ref.~\cite{Katsufuji_PRL'96}. } \label{fig:alpha}
\end{figure}

We now consider the other limit $\omega\sim\gamma_{ab}$.  The experimental measurements of $\gamma_{ab}$ in the normal state at $T=9$~K by Katsufuji et al.~\cite{Katsufuji_PRL'96} indicate that the quasiparticle scattering rate at the optical frequency $\omega \sim 0.8$~eV is $\gamma_{ab}\approx 0.4$~eV.  Assuming that the magnitude of the quasiparticle scattering rate is the same in the superconducting state at $T<T_c$, the real and imaginary parts of $\alpha(\omega)$ become of the same order as shown in Fig.~\ref{fig:alpha}. Then, according to Eq.~\eqref{eq:thetaK1}, both real and imaginary parts of the Hall conductivity \eqref{eq:estimates_high-3} give comparable contributions to the Kerr angle.  Using Eq.~\eqref{epsilon_ab} with the value $\gamma_{ab}\approx 0.4$~eV quoted above and Eqs.~\eqref{alpha}, \eqref{eq:thetaK1}, and \eqref{eq:estimates_high-3}, we estimate the Kerr angle in this case as
  \begin{align}
  \theta_K \approx 43 \: \rm nrad.
  \label{estimate_kerr}
  \end{align}
This estimate is of the same order of magnitude as the experimentally observed $\theta^{\rm exp}_K \approx 65$~nrad~\cite{xia_prl'06}.

In the process of calculating the estimates, we made numerous approximations and assumptions.  The values of many parameters that appear in the theory are unknown, and more measurements are needed to determine them.  Nevertheless, the final estimate \eqref{estimate_kerr} is encouraging and indicates that impurity scattering may provide a viable explanation for the polar Kerr effect in $\rm Sr_2RuO_4$.

When $\alpha'$ and $\alpha''$ in Eq.~\eqref{eq:thetaK1} are of the same order, $\theta_K$ is dominated by $\sigma''_{xy}$, which is much greater than $\sigma'_{xy}$.  Thus, $\theta_K(T)\propto\sigma''_{xy}(T)$, so the temperature dependence of $\sigma''_{xy}(T)$ determines the temperature dependence of $\theta_K(T)$, assuming that $\alpha$ at high frequencies does not depend significantly on temperature for $T<T_c$.  Then, Eq.~\eqref{eq:sigma3T} and the plot in Fig.~\ref{fig:kerr-T} give the temperature dependence of the normalized Kerr angle $\theta_K(T)/\theta_K(0)$.  This theoretical result can be directly compared with the temperature dependence of $\theta_K(T)$ experimentally measured in Ref.~\cite{xia_prl'06}.

\section{Conclusions} \label{sec:conclusions}

In this paper, we calculate the anomalous (spontaneous) ac Hall conductivity $\sigma_{xy}(\omega)$ for a chiral $p_x + ip_y$ superconductor, such as $\rm Sr_2RuO_4$, in the long wavelength limit $\bm q\rightarrow 0$ in the absence of an external magnetic field.  We show that the anomalous Hall conductivity vanishes for a translationally-invariant system, and a non-zero result requires presence of impurities.  Non-zero contributions to the anomalous Hall conductivity appear in the higher order (above the second order) of the perturbation theory in impurity scattering.  We consider two models of disorder potential: Gaussian and non-Gaussian.  The Gaussian model is characterized by the second moment of the random impurity potential, whereas the non-Gaussian model has a non-zero third moment.  For the Gaussian disorder model, we present a symmetry argument and demonstrate by direct calculations that a non-zero anomalous Hall conductivity requires particle-hole asymmetry of the electron spectrum.  Thus, the magnitude and the sign of $\sigma_{xy}(\omega)$ depend on the band-structure parameter $N'(0)$, which is the derivative of the normal density of states at the Fermi level.  On the other hand, the anomalous Hall conductivity in the non-Gaussian model is not proportional to the particle-hole asymmetry parameter, and its sign is determined by the sign of the impurity potential.  Therefore, in principle, these two models can be distinguished experimentally by introducing positively and negatively charged impurities.  However, in practice, such a discrimination may be difficult, because the sign of the anomalous Hall conductivity also depends on the sign of the chirality of the superconducting $p_x \pm ip_y$ order parameter.

By calculating the lowest-order non-vanishing Feynman diagrams shown in Figs.~\ref{fig:skew_diagrams} and \ref{fig:2nd_order_line}, we obtain closed-form expressions for the frequency and temperature dependences of the anomalous Hall conductivities originating from these two models.  As a function of frequency, the calculated Hall conductivities $\sigma_{xy}(\omega)$ have a finite real value at $\omega=0$ and exhibit singularities at the threshold of photon absorption $\omega=2\Delta_0$.  At high frequencies $\omega \gg \Delta_0$, the dominant contribution to the anomalous Hall conductivity comes from the imaginary part $\sigma_{xy}''(\omega)$ which decays as $1/\omega^3$, see  Figs.~\ref{fig:FOm}, \ref{fig:S}, and \ref{fig:G}.  As a function of temperature $T$, the high-frequency $\sigma_{xy}''(\omega,T)$ increases linearly with the decrease of temperature near $T_c$ and saturates at $T\to0$, as shown in Fig.~\ref{fig:kerr-T}.

In order to estimate whether the Gaussian or non-Gaussian term gives a dominant contribution to the anomalous Hall conductivity for $\rm Sr_2RuO_4$, it is necessary to know the parameters characterizing the electron spectrum and the strength of the disorder.  The latter are poorly known.  Nevertheless, by making a number of assumptions and approximations, we obtain numerical estimates, which show that the non-Gaussian term dominates over the Gaussian one both at high and low frequencies.

The polar Kerr angle $\theta_K$ is proportional to the ac Hall conductivity $\sigma_{xy}(\omega)$.  However, the proportionality relation also involves the complex refraction coefficient $n(\omega)$, which is poorly known for $\rm Sr_2RuO_4$.  Using a Drude model with the experimentally estimated parameters to obtain $n(\omega)$ and the anomalous Hall conductivity from our calculations, we estimate the magnitude of the polar Kerr angle at $\omega=0.8$~eV as $\theta_K\approx 43$ nrad, which is of the same order as the experimentally observed value of 65 nrad \cite{xia_prl'06}.  Despite numerous assumptions and approximations used, this result is encouraging and indicates that impurity scattering models may provide a viable explanation for the polar Kerr effect in $\rm Sr_2RuO_4$.  However, the main conclusion of our paper is not a particular value of $\theta_K$, but the qualitative understanding that the magnitude of the anomalous Hall conductivity and the Kerr angle are directly proportional to some power of the impurity concentration and should strongly vary among different samples.  Systematic measurements of the Kerr angle as a function of concentration of the intentionally introduced impurities would be very desirable.  The study should be performed for sufficiently low concentration in the range $\Delta_0\gg 1/\tau$, where impurities do not affect the superconducting order significantly, but do affect the polar Kerr effect.

\begin{acknowledgments}
The authors (RML and VMY) acknowledge hospitality of the Kavli Institute for Theoretical Physics, Santa Barbara (supported by the National Science
Foundation under Grant No.~PHY05-51164), where a part of this work was
done during the program ``Low Dimensional Electron Systems''.  We would like to thank A.~Kapitulnik, C.~Kallin, S.~Das~Sarma, H.~D.~Drew, V.~Galitski, E.~Hwang, and E.~Rossi for stimulating discussions. This work is supported by JQI-PFC-NSF. RML is also partially supported by DARPA-QuEST-AFOSR.
\end{acknowledgments}

\appendix

\section{The Peierls-Onsager substitution for superconductors}
\label{App:substitution}

The BCS mean-field action for a superconducting system is given in
Eq.~\eqref{S_el}. The corresponding Hamiltonian $\hat H(\bm p)$ has
the form of a $2\times2$ Nambu matrix
\begin{equation}
  \hat H(\bm p)=\left(
  \begin{array}{cc}
  \xi(\bm p) & \Delta(\bm p) \\
  \Delta^*(\bm p) & -\xi(-\bm p)
  \end{array}  \right),
\label{H(p)}
\end{equation}
acting on the spinor $[\psi(\bm p),\psi^\dag(-\bm p)]$.  We omitted
the spin indices of the fermion operators, because they are not
essential for the consideration here.  The notation in
Eq.~\eqref{H(p)} is the same as in Sec.~\ref{Sec:Hamiltonian}.  For
simplicity, we will assume that $\xi(-\bm p)=\xi(\bm p)$, which is
the case for most centrosymmetric materials, including $\rm
Sr_2RuO_4$.

When the system is subject to an electromagnetic field, the question
arises how the vector potential $\bm A$ should be introduced into
the Hamiltonian \eqref{H(p)}.  For non-superconducting systems, it
is introduced via the Peierls-Onsager substitution $\bm p\to\bm
p-e\bm A$, and it is tempting to make the same substitution in
Eq.~\eqref{H(p)}.  However, one should be careful and re-examine how
this substitution originates from the requirement of gauge
invariance.  The two diagonal terms in Eq.~\eqref{H(p)} give the
following contributions $\int d^3r\,\psi^\dag(\bm
r)\,\xi(-i\bm\nabla)\,\psi(\bm r)$ and $\int d^3r\,\psi(\bm
r)\,\xi(-i\bm\nabla)\,\psi^\dag(\bm r)$ to the Hamiltonian of the
system in the real-space representation.  When we make a gauge
transformation of the fermion operators $\psi(\bm r)\to\psi(\bm
r)\,e^{i\varphi(\bm r)}$, the gradients of the phase $\varphi$ need
to be compensated by the gauge transformation of the vector
potential $\bm A$.  This leads to the substitution $\bm p\to\bm
p-e\bm A$ and $\bm p\to\bm p+e\bm A$ in the upper and lower diagonal
term in Eq.~\eqref{H(p)}.  The substitutions are different because
of the different order of $\psi$ and $\psi^\dag$ in these terms.
This is very well known \cite{Lee_RMP'06}.

For unconventional superconductors, where the pairing potential
$\Delta(\bm p)$ explicitly depends on the electron momentum $\bm p$,
the question arises whether the Peierls-Onsager substitution should
be made in $\Delta(\bm p)$.  However, it is not clear whether the
rule $\bm p\to\bm p-e\bm A$ or $\bm p\to\bm p+e\bm A$ should be used
in the off-diagonal term in Eq.~\eqref{H(p)}.  Actually, the
Peierls-Onsager substitution is not needed in the off-diagonal term,
because, when properly written, it is already gauge-invariant
without introduction of the vector potential $\bm A$.  When the
order parameter varies in space, the pairing potential
\eqref{Delta_xy} should be written as the symmetrized combination,
i.e., the anticommutator, of the two-component order parameter
$\bm\Psi=(\Delta_x,i\Delta_y)$ and the momentum operator $\bm
p=(p_x,p_y)$.  The BCS Hamiltonian \eqref{H(p)} acquires the
following form in the real space
\begin{equation}
  \hat H=\left(
  \begin{array}{cc}
  \xi(-i\bm\nabla-e\bm A)
  & -i(\bm\nabla\cdot\bm\Psi+\bm\Psi\cdot\bm\nabla)/2 \\
  -i(\bm\nabla\cdot\bm\Psi^*+\bm\Psi^*\cdot\bm\nabla)/2
  & -\xi(-i\bm\nabla+e\bm A)
  \end{array}  \right).
\label{Psi}
\end{equation}
It is now easy to check that the off-diagonal terms in
Eq.~\eqref{Psi} are gauge-invariant under the simultaneous phase
transformation of the fermion operators $\psi(\bm r)\to\psi(\bm
r)\,e^{i\varphi(\bm r)}$, $\psi^\dag(\bm r)\to\psi^\dag(\bm
r)\,e^{-i\varphi(\bm r)}$ and the superconducting order parameter
$\bm\Psi(\bm r)\to\bm\Psi(\bm r)\,e^{2i\varphi(\bm r)}$.  Notice
that the gradient operators in Eq.~\eqref{Psi} act to the right on
both $\psi(\bm r)$ and $\bm\Psi(\bm r)$.

Thus, it is incorrect to make the Peierls-Onsager substitution $\bm
p\to\bm p-e\bm A$ in the pairing potential $\Delta(\bm p)$, as it
was proposed in Refs.~\cite{Balatskii_JETP'85,mineev_prb'08}.  Even
if such a substitution were made, it would have generated the term
$\bm A\cdot\bm\Psi^*\int d^3p\,\psi(-\bm p)\,\psi(\bm p)$ and its
Hermitian conjugate.  These terms vanish, because they reverse sign
upon commutation of the fermion operators and changing sign of the
variable of integration $\bm p$.   This conclusion also holds when
the spin indices of the fermion operators $\psi$ are restored,
because the spin-triplet order parameter $\bm\Psi$ is a symmetric
spin tensor, so there is no sign change upon exchange of the spin
indices.  Curiously, when the conventional vertex of interaction
with the electromagnetic field is dressed by the impurity line as
shown in Fig.~\ref{fig:single_line}b, the resulting vertices
$\hat\Lambda_x$ \eqref{Lambda_x} and $\hat\Lambda_y$
\eqref{Lambda_y} have the structure similar to the terms discussed
above.  However, these vertices do not vanish upon fermion
commutation, because they change sign upon exchange of the fermion
frequencies $\varpi_l$ and $\varpi_l+\omega_n$.

The BCS Hamiltonian in the form \eqref{Psi}, with the pairing
potential written as the anticommutator, was introduced and
discussed in Ref.~\cite{Volovik_JETP'88} and, more recently, in
Ref.~\cite{Read_PRB'09}.  This form was used in most papers on the
subject, e.g., in Ref.~\cite{yakovenko_prl'07}, and was recently
recognized in Ref.~\cite{mineev'09}. For a chiral $d+id$
superconductor, a microscopic derivation from a lattice model was
given in Ref.~\cite{Vafek_PRB'01}.

The current operator is obtained by expanding the Hamiltonian
\eqref{Psi} to the first order in $\bm A$, which gives the same
result as in Eq.~\eqref{S_em}.  Notice that the current operator for
a superconducting systems is not given by the derivative
$\partial\hat H/\partial\bm p$ of the Hamiltonian \eqref{H(p)} with
respect to the momentum $\bm p$, because $\bm p$ and $\bm A$ do not
appear in the combination $\bm p-e\bm A$.  For this reason, in
contrast to non-superconducting systems, the anomalous Hall
conductivity of a chiral superconductor is not expressed in terms of
the Berry curvature, as discussed in the next Appendix.

\section{Chiral superconductors vs.\ the TRSB topological insulators and metals}
\label{Sec:Insulators}

Insulators with a non-trivial band topology (the topological
insulators) can be divided into two classes: the
time-reversal-invariant and the time-reversal-symmetry-breaking (the
TRSB topological insulators) \cite{Kane_PRL'05, Moore_PRB'07,
Roy_arxiv'06, Qi_PRB'08}.  Here we discuss similarities and
differences between chiral superconductors and the TRSB topological
insulators.  Some of their features, such as the energy gap in the
single-particle spectrum and the presence of chiral edge states, are
common.  However, the electromagnetic properties of chiral
superconductors are different from those of topological insulators,
e.g., because superconductors are not insulators.

Following the pioneering work by Haldane~\cite{Haldane_PRL'88}, many
authors investigated periodic systems with a topologically
non-trivial band structure resulting from a distribution of the
effective Aharonov-Bohm fluxes inside a unit cell.  In such systems,
the time-reversal symmetry may be broken even when the total flux
through the unit cell is zero.  Typically, such systems have a
non-zero Berry curvature and exhibit anomalous Hall effect.  A model
of this kind, involving formation of the $d_{xy}+id_{x^2-y^2}$
density wave, was proposed in Ref.~\cite{Tewari_PRL'08} to explain
the Kerr effect in underdoped cuprates~\cite{Xia_PRL'08}.  The
anomalous Hall effect was studied for this model earlier in
Ref.~\cite{Yakovenko_PRL'90}, and the anomalous Nernst effect was
discussed in Ref.~\cite{Tewari_PRB'08}.

However, it is important to emphasize that the origin of the
anomalous Hall effect in these models has nothing to do with
superconductivity.  Thus, these models are not relevant for $\rm
Sr_2RuO_4$, where the Kerr effect appears at the superconducting
$T_c$ \cite{xia_prl'06}.  If the Kerr effect in $\rm Sr_2RuO_4$ were
due to a topologically non-trivial band structure, it would have
been visible above the superconducting $T_c$, as in the underdoped
cuprates \cite{Xia_PRL'08}.  Thus, for the applications to $\rm
Sr_2RuO_4$, the question is whether a non-zero anomalous Hall
conductivity and the Kerr effect may originate from the TRSB solely
due to chiral superconductivity and not due to a topologically
non-trivial band structure.  For this purpose, one may consider the
simple one-band parabolic dispersion law in the $(x,y)$ plane and
ignore complications of the real band structure of $\rm Sr_2RuO_4$,
which consists of three distinct
sheets~\cite{Mackenzie_RevModPhys'03,maeno_PhysToday'01}.  It was
shown in Ref.~\cite{lutchyn_prb'08} that the anomalous Hall
conductivity of a clean $p_x+ip_y$ superconductor vanishes even when
an arbitrary electron dispersion is considered.

For non-superconducting systems, the current operator can be
expressed in terms of the derivatives $\partial\hat H/\partial\bm p$
of the Hamiltonian $\hat H$ with respect to the electron
quasi-momenta $\bm p$.  For a topologically non-trivial band
structure, the off-diagonal matrix elements of the current operator
between different bands produce a non-zero Berry curvature, and
the anomalous Hall conductivity can be expressed completely in terms
of the Berry
curvature~\cite{Haldane_PRL'88,Yakovenko_PRL'90,Haldane_PRL'04,Tewari_PRL'08,Tewari_PRB'08,Sinitsyn_JPCM'08}.

The situation is different for chiral superconductors.  As discussed
in Appendix \ref{App:substitution}, the current operator for a
superconductor cannot be expressed in terms of the derivatives
$\partial\hat H/\partial\bm p$ of the BCS Hamiltonian and is not
related to the group velocity $\partial E(\bm p)/\partial\bm p$ of
the Bogolyubov quasiparticles, as discussed around Eq.~(3) in
Ref.~\cite{Lee_RMP'06}.  Thus, the Hall conductivity of a
superconductor cannot be expressed in terms of the Berry curvature;
moreover, the Hall conductivity vanishes, as discussed in
Sec.~\ref{Sec:Theories}.  In contrast to the TRSB topological
insulators, the effective action of a chiral superconductor contains
only one part of the Chern-Simons
term~\cite{Volovik_JETP'88,Goryo_PhysLettA'98,Goryo_PhysLettA'99,Horovitz_epl'02,Horovitz_PRB'03,Stone_PRB'04,yakovenko_prl'07,lutchyn_prb'08},
involving $A_0$ and $A_x$ or $A_y$, but does not contain the other
part, involving $A_x$ and $A_y$.  The reason is that the
superconducting order parameter $\Delta$ has the phase $\Phi$, which
appears in the effective action and modifies requirements imposed by
the gauge symmetry.  When the self-consistent dynamics of the phase
$\Phi$ is taken into account, the anomalous Hall conductivity
vanishes for a clean chiral
superconductor~\cite{Goryo_PhysLettA'98,Goryo_PhysLettA'99,Horovitz_epl'02,Horovitz_PRB'03,lutchyn_prb'08,roy_prb'08}.

\section{Calculation of the function \boldmath $H(\omega_n)$}
\label{App:eval_I}

In the following Appendixes, we denote the gap $\Delta_0$ by simply $\Delta$ in order to shorten lengthy mathematical expressions.

\begin{figure} \centering
\includegraphics[width=0.75\linewidth]{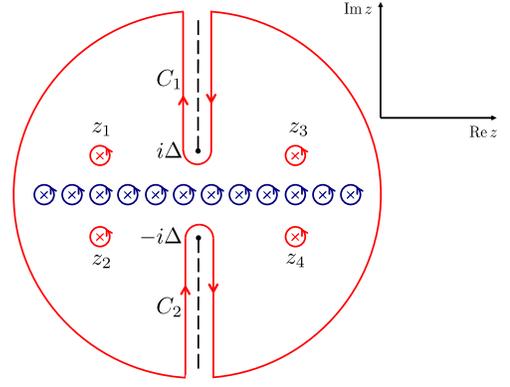}
  \caption{(Color online) Contours of integration chosen to evaluate the Matsubara sums in Eqs.~(\ref{eq:Isum}) and (\ref{eq:K2222}).  The initial contour (blue lines) encircles the series of points along the horizontal axis, where $\tan(z/2T)$ has pole singularities.  Then, the contour is deformed to infinity, encircling the poles $z_j$ and the brunch cuts $C_1$ and $C_2$ of the function under the integral (red lines). }
  \label{fig:contourA}
\end{figure}

Here we calculate the function $H(\omega_n)$ in Eq.~\eqref{trip}:
  \begin{align}
  &\! H(\omega_n) = T \Delta \sum_l
  \left(\frac{\varpi_l+\omega_n}{\sqrt{\Delta^2+(\varpi_l+\omega_n)^2}}
  -\frac{\varpi_l}{\sqrt{\Delta^2+\varpi_l^2}}\right)
  \nonumber \\
  \!&\! \times \left(\frac{1}{\sqrt{\Delta^2+\varpi_l^2}}
  -\frac{1}{\sqrt{\Delta^2+(\varpi_l+\omega_n)^2}}\right)^2
  \label{H1} \\
  \!& = T \Delta \sum_l\left[\frac{\varpi_l+\omega_n}
  {[(\varpi_l+\omega_n)^2+\Delta^2]^{3/2}}
  -\frac{\varpi_l}{(\varpi_l^2+\Delta^2)^{3/2}} \right.
  \label{H2} \\
  \!\!&\! \left. \!+\! \frac{3\varpi_l+\omega_n}
  {(\varpi_l^2\!+\!\Delta^2)\sqrt{(\varpi_l\!+\!\omega_n)^2\!+\!\Delta^2}}
  \!-\! \frac{3\varpi_l+2\omega_n}{[(\varpi_l\!+\!\omega_n)^2
  \!+\!\Delta^2]\sqrt{\varpi_l^2\!+\!\Delta^2}} \! \right] \!.
  \nonumber
  \end{align}
In going from Eq.~\eqref{H1} to \eqref{H2}, the terms odd in $\varpi_l$ and $\varpi_l+\omega_n$ vanish after summation over $l$.  After the variable shift $\varpi_l\to \varpi_l - \omega_n$ in the first and third terms in Eq.~\eqref{H2}, $H(\omega_n)$ can be written as
  \begin{align}
  H(\omega_n) & = T \Delta \sum_l f(\varpi_l),
  \label{eq:Isum} \\
  f(\varpi_l) & = \frac{3\varpi_l-2\omega_n}{\sqrt{\varpi_l^2+\Delta^2}
  [(\varpi_l - \omega_n)^2 + \Delta^2]}
  \nonumber \\
  & - \frac{3\varpi_l+2\omega_n}{\sqrt{\varpi_l^2+\Delta^2}
  [(\varpi_l + \omega_n)^2 + \Delta^2]} .
  \end{align}
The Matsubara sum in Eq.~\eqref{eq:Isum} can be replaced by integration in the complex plane
  \begin{align}  \label{complex}
  T \sum_l f(\varpi_l) = \frac12 \int f(z) \tan\left(\frac{z}{2T}\right)
  \frac{dz}{2\pi i},
  \end{align}
with the contour of integration being a series of circles around the points on the horizontal axis in Fig.~\ref{fig:contourA}, where the function $\tan(z/2T)$ has pole singularities.  Then, this contour is expanded to a circle of an infinite radius and the contours encircling the poles $z_j$ and the brunch cuts of the function $f(z)$, as shown in Fig.~\ref{fig:contourA}
  \begin{align}
  T \sum_l f(\varpi_l) & = \frac12 \sum_{j=1,2,3,4}
  \tan\left(\frac{z_j}{2T}\right) \mathrm{Res} f(z_j)
  \nonumber \\
  & + \frac12 \int_{C_1,C_2} f(z) \tan\left(\frac{z}{2T}\right)
  \frac{dz}{2\pi i}.
  \label{eq:Mats_sum}
  \end{align}
The contribution from the poles at $z_{1,2}=-\omega_n \pm i\Delta$ and $z_{3,4}=\omega_n \pm i\Delta$ is
  \begin{align} \label{App:I1}
  \frac12 \sum_{j=1,2,3,4} & \tan\left(\frac{z_j}{2T}\right)
  \mathrm{Res} f(z_j)=\frac{1}{2\Delta} \tanh\left(\frac{\Delta}{2T}\right)
  \\
  & \times \left(\frac{\omega_n+3i\Delta}{\sqrt{\omega_n(\omega_n+2i\Delta)}}
  +\frac{\omega_n-3i\Delta}{\sqrt{\omega_n(\omega_n-2i\Delta)}}\right).
  \nonumber
  \end{align}
The contribution from the branch cuts $C_1$ and $C_2$ is
  \begin{align} \label{App:I2}
  & \frac12 \int_{C_1,C_2} f(z) \tan\!\left(\frac{z}{2T}\right)
  \frac{dz}{2\pi i} = \frac{1}{\pi} \int^\infty_\Delta
  \frac{dx}{\sqrt{x^2-\Delta^2}}
  \\
  & \times \left(\frac{3ix-2\omega_n}{\Delta^2+(ix-\omega_n)^2}
  -\frac{3ix+2\omega_n}{\Delta^2+(ix+\omega_n)^2}\right)
  \tanh\left(\frac{x}{2T}\right).
  \nonumber
  \end{align}
Combining the two contributions \eqref{App:I1} and \eqref{App:I2} to Eq.~\eqref{eq:Isum}, we arrive to Eq.~(\ref{eq:Ifinal}) for $H(\omega_n)$.

\section{Calculation of the dressed vertex \boldmath $\hat\Gamma_j(\bm p)$}
\label{vertex_corr}

Here we calculate the dressed vertex $\hat\Gamma_j(\bm p)$ by solving the Bethe-Salpeter equation shown in Fig.~\ref{fig:self_vertex}b and given analytically in Eq.~\eqref{Bethe-Salpeter}.  According to Eq.~\eqref{eq:dressed_bubble}, to calculate anomalous Hall response, we need to obtain the dressed vertex function $\hat \Gamma_y$. By iterating Eq.~\eqref{Bethe-Salpeter}, we find a geometric series for $\hat \Gamma_y$
  \begin{align} \label{iterations}
  & \hat\Gamma_y(\bm p) = \hat\gamma_y(\bm p) + \hat M_y(\varpi_l,\omega_n)
  \\
  & + n_iu_0^2 \sum_{\bm p'} \hat\tau_3 \hat{\cal G}_{\bm p'}(\varpi_l)
  \hat M_y(\varpi_l,\omega_n) \hat{\cal G}_{p'}(\varpi_l + \omega_n) \hat\tau_3
  + \ldots \, ,
  \nonumber
  \end{align}
  where the function $\hat M_y$ is
  \begin{align}
  \hat M_y(\varpi_l,\omega_n) & = n_iu_0^2 \sum_{\bm p'}
  \hat\tau_3 \hat{\cal G}_{\bm p'}(\varpi_l) \hat\gamma_y(\bm p)
  \hat{\cal G}_{p'}(\varpi_l + \omega_n) \hat\tau_3
  \nonumber \\
  & = \frac{\Delta_y p_F v_F}{2\pi\tau} L(\varpi_l,\omega_n)
  (-i a_1 \hat\tau_1+ a_2 \hat\tau_2).
  \end{align}
The functions $a_1$, $a_2$, and $L$ are defined in Eqs.~(\ref{eq:a1}), (\ref{eq:a2}), and (\ref{eq:L}), respectively.  The next-order term in Eq.~\eqref{iterations} is
  \begin{align}
  & n_iu_0^2 \sum_{\bm p'} \hat\tau_3 \hat{\cal G}_{\bm p'}(\varpi_l)
  \hat M_y(\varpi_l,\omega_n) \hat{\cal G}_{p'}(\varpi_l+\omega_n) \hat\tau_3
  \nonumber \\
  & =  \frac{\Delta_y p_F v_F}{2\pi\tau} L(\varpi_l,\omega_n) \,
  (-i a_1 \hat\tau_1+ a_2 \hat\tau_2) \, (b_1\hat\tau_0+i b_2\hat\tau_3),
  \nonumber
  \end{align}
where the functions $b_1$ and $b_2$ are defined in Eqs.~(\ref{eq:b1}) and~(\ref{eq:b2}).  We observe that the higher-order terms bring the powers of the factor $b_1\hat\tau_0+i b_2\hat\tau_3$, so we need to sum the following geometrical series
  \begin{align} \label{geometrical}
  \sum_{k=0}^{\infty} (b_1\hat\tau_0+i b_2\hat\tau_3)^k \! = \!
  \frac{1-b_1}{(1-b_1)^2+b^2_2}\hat\tau_0 + \frac{ib_2}{(1-b_1)^2+b^2_2}\hat\tau_3.
  \end{align}
Using this result in Eq.~\eqref{iterations}, we find the dressed vertex $\hat\Gamma_y(\bm p)$
  \begin{align}
  \hat\Gamma_y(\bm p) & = \gamma_y(\bm p) + \frac{\Delta_y p_F v_F}{2\pi\tau}
  L(\varpi_l,\omega_n) (-i a_1\hat\tau_1 + a_2\hat\tau_2)
  \nonumber \\
  & \times \left(\frac{1-b_1}{(1-b_1)^2+b^2_2} \hat\tau_0
  + \frac{i b_2}{(1-b_1)^2+b^2_2} \hat\tau_3 \right).
  \label{Gamma}
  \end{align}

\section{Calculation of the function \boldmath $K(\omega_n)$}
\label{app:calc_K}

\begin{figure} \centering
\includegraphics[width=0.75\linewidth]{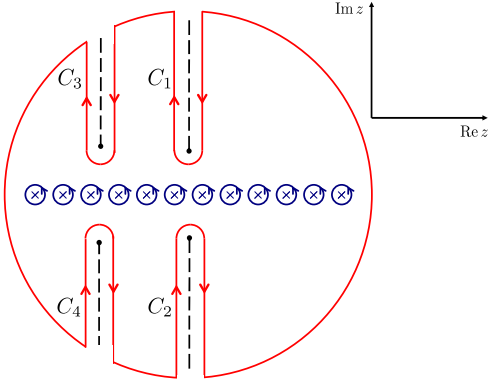}
  \caption{(Color online) Contour of integration chosen to evaluate the Matsubara sum in Eqs.~(\ref{eq:K111}) and (\ref{eq:contourB}).  The initial contour (blue lines) encircles the series of points along the horizontal axis, where $\tan(z/2T)$ has pole singularities.  Then, the contour is deformed to infinity and goes along the brunch cuts $C_{1,2,3,4}$ of the function under the integral (red lines). } \label{fig:contourB}
\end{figure}

The expression for the function $K(\omega_n)$ defined in Eq.~(\ref{eq:K0}) can be written as a sum of two terms
  \begin{align}
  K(\omega_n) & = K_1(\omega_n) + K_2(\omega_n),
  \label{K-sum} \\
  K_1(\omega_n) & = \sum_l\! \frac{T\Delta}{2\varpi_l\!+\!\omega_n} \!
  \left[ \frac{3}{\sqrt{(\varpi_l\!+\!\omega_n)^2\!+\!\Delta^2}}
  \!-\! \frac{3}{\sqrt{\varpi_l^2\!+\!\Delta^2}} \right] \! ,
  \label{eq:K111} \\
  K_2(\omega_n) & = \sum_l \! \frac{T\Delta}{2\varpi_l+\omega_n}
  \label{eq:K222} \\
  & \times \left( \frac{\sqrt{(\varpi_l+\omega_n)^2+\Delta^2}}{\varpi_l^2+\Delta^2}
  - \frac{\sqrt{\varpi_l^2+\Delta^2}}{(\varpi_l+\omega_n)^2+\Delta^2} \right).
  \nonumber
  \end{align}
One can notice that the above expressions for $K_1(\omega_n)$ and $K_2(\omega_n)$ do not have a singularity at $\varpi_l=-\omega_n/2$.  Similarly to Eq.~\eqref{eq:Isum}, the Matsubara sum in Eq.~(\ref{eq:K111}) can be replaced by integration in the complex plane around the series of points along the horizontal axis in Fig.~\ref{fig:contourB}, where the function $\tan(z/2T)$ has pole singularities.  The contour of integration is then deformed to go along the branch cuts $C_{1,2,3,4}$ of the function under the integral, as shown in Fig.~\ref{fig:contourB}.  The result is
  \begin{align} \label{eq:K22}
  K_1(\omega_n) = - 12 \, \tilde\omega_n \int_1^\infty \frac{dx}{2\pi}
  \frac{\tanh(x\Delta/2T)}{\sqrt{x^2-1} \, (\tilde\omega_n^2+4x^2)},
  \end{align}
where $\tilde\omega_n=\omega_n/\Delta$.

To evaluate the sum in Eq.~(\ref{eq:K222}), we first shift the frequency $\varpi_l\to\varpi_l+\omega_n$ in the first term to obtain:
  \begin{align}
  K_2(\omega_n) = T \Delta & \sum_l \left[
  \frac{\sqrt{\varpi_l^2+\Delta^2}}{(2\varpi_l-\omega_n)
  ([\varpi_l-\omega_n]^2+\Delta^2)} \right.
  \nonumber \\
  & \left. - \frac{\sqrt{\varpi_l^2+\Delta^2}}{(2\varpi_l+\omega_n)
  ([\varpi_l+\omega_n]^2+\Delta^2)} \right].
  \label{eq:K2222}
\end{align}
The contour of integration used to evaluate the sum in Eq.~(\ref{eq:K2222}) is shown in Fig.~\ref{fig:contourA}.  The poles yield
  \begin{align}  \label{eq:K11}
  K_{2a}(\omega_n) & = \frac12 \tanh\left(\frac{\Delta}{2T}\right)
  \\
  & \times \left(\frac{\sqrt{1-(1-i\tilde\omega_n)^2}}{2i+\tilde\omega_n}
  - \frac{\sqrt{1-(1+i\tilde\omega_n)^2}}{2i-\tilde\omega_n}\right),
  \nonumber
  \end{align}
and the branch cuts contribute
  \begin{align}  \label{eq:K33}
  & K_{2b}(\omega_n) = -2 \int_1^\infty \frac{dx}{2\pi} \sqrt{x^2-1}
  \tanh\left(\frac{\Delta x}{2T}\right)
  \\
  \!&\! \times \!\!\left[\!
  \frac{1}{(2ix\!-\!\tilde\omega_n)(1-[x\!+\!i\tilde\omega_n]^2)}
  \!-\!\frac{1}{(2ix\!+\!\tilde\omega_n)(1-[x\!-\!i\tilde\omega_n]^2)}
  \! \right] \! .
  \nonumber
  \end{align}
Eqs.~(\ref{eq:K22}), (\ref{eq:K11}), and (\ref{eq:K33}) give the expression for $K(\omega_n)$ shown in Eq.~(\ref{eq:K}).

\section{Calculation of the function \boldmath $J(\omega_n)$}
\label{app:calc_J}

The Matsubara sum \eqref{eq:J0} for the function $J(\omega_n)$ can be separated into the terms with poles and branch cuts
  \begin{align}  \label{eq:J_A1}
  J(\omega_n) & = T \Delta \sum_l \left( \frac{1}{\varpi_l^2+\Delta^2}
  + \frac{1}{(\varpi_l+\omega_n)^2+\Delta^2} \right.
  \nonumber \\
  & \left. - \frac{2}
  {\sqrt{(\varpi_l+\omega_n)^2+\Delta^2} \sqrt{\varpi_l^2+\Delta^2}} \right).
  \end{align}
After shifting $\varpi_l$ in the second term of Eq.~(\ref{eq:J_A1}), we get  \begin{align} \label{J-pole+cut}
  J(\omega_n) = \sum_l \left(\frac{2T\Delta}{\varpi_l^2\!+\!\Delta^2} -
  \frac{2T\Delta}{\sqrt{(\varpi_l\!+\!\omega_n)^2\!+\!\Delta^2}
  \sqrt{\varpi_l^2\!+\!\Delta^2}}\right).
  \end{align}
The first term in Eq.~\eqref{J-pole+cut} has only simple poles
  \begin{align} \label{J-poles}
  \sum_l \frac{2T \Delta}{\varpi_l^2+\Delta^2}
  = \tanh\left(\frac{\Delta}{2T}\right).
  \end{align}
The contribution from the second term in Eq.~\eqref{J-pole+cut} can be calculated by using the contour shown in Fig.~\ref{fig:contourB}
  \begin{align}
  & \sum_{\varpi_l} \frac{2T\Delta}{\sqrt{(\varpi_l\!+\!\omega_n)^2\!+\!\Delta^2}
  \sqrt{\varpi_l^2\!+\!\Delta^2}}
  = \frac{2}{\pi} \int\limits^\infty_\Delta dx \,
  \frac{\tanh\left(\frac{x}{2T}\right)}{\sqrt{x^2-\Delta^2}}
  \nonumber\\
  & \times \left( \frac{\Delta}{{\sqrt{(ix+\omega_n)^2+\Delta^2}}}
  +\frac{\Delta}{{\sqrt{(ix-\omega_n)^2+\Delta^2}}} \right) .
  \label{eq:contourB}
  \end{align}
By combining the two contributions \eqref{J-poles} and \eqref{eq:contourB}, we obtain the result shown in Eq.~(\ref{eq:J}).


\begin{thebibliography}{48}
\bibitem{edges} For a system with edges, the general argument using translational symmetry may be not directly applicable.  However, the optical measurements \cite{xia_prl'06} were performed in a contactless manner using a small laser beam far away from the sample edges.  Thus, edge effects are not relevant and are not considered in our paper.
\bibitem{footnote2}
It can be transformed into pairing with parallel spins by changing the spin quantization axis.
\bibitem{footnote3}
It is easy to check that a similar calculation for $Q_{yx}(\omega_n)$ gives $Q_{yx}(\omega_n)=-Q_{xy}(\omega_n)$, so it indeed represents the Hall response.
\bibitem{footnote:theta(T)}
As discussed in Sec.~\ref{Sec:Kerr-estimate}, because $\theta_K$ is proportional to the dominant part of $\sigma_{xy}$, Eq.~\eqref{eq:sigma3T} also gives the temperature dependence of the normalized Kerr angle $\theta_K(T)/\theta_K(0)$.
\bibitem{footnote_Goryo}
Using the estimates for $\tau$ and $n_i$ quoted in the text, we find that
the expansion parameter $u_0 N(0) \sim 1$ has the borderline value for applicability of the perturbation theory with respect to the impurity potential.
\bibitem{footnote4}
We estimate $\omega_{ab}$ using the formula
$\omega_{ab}=\omega_c\sqrt{N^{ab}_{\rm eff}/N^{c}_{\rm eff}}$, where
$\omega_c=0.32$ eV is the out-of-plane plasma frequency, and
$N^{ab}_{\rm eff}/N^{c}_{\rm eff}\sim 80$ is the ratio of the effective densities of carriers along the different axes, as explained in Ref.~\cite{Katsufuji_PRL'96}.
\end{thebibliography}

\end{document}